\documentclass[12pt]{iopart}

\usepackage{amsbsy}
\usepackage{amssymb}
\usepackage{graphicx}
\usepackage{xcolor}
\usepackage{hyperref}
\usepackage{array}
\newcolumntype{L}{>{$}l<{$}} 
\usepackage{cite}
\usepackage{cleveref}

\newcommand{\symm}{\scriptsize\textrm{h}}
\newcommand{\sech}{\textrm{sech}}
\def\de{\mathrm{d}}

\begin{document}

\title{Optimal resetting strategies for search processes in heterogeneous environments}

\author{Gregorio García-Valladares$^{1}$, Carlos A. Plata$^{1}$, Antonio Prados$^{1}$, and Alessandro Manacorda$^{2,*}$ }
\address{$^{1}$Física Teórica, Universidad de Sevilla, Apartado de Correos 1065, E-41080 Sevilla, Spain}
\address{$^{2}$Department of Physics and Materials Science, University of Luxembourg, L-1511 Luxembourg}
\address{$^{*}$Authors to whom any correspondence should be addressed.}	
\ead{alessandro.manacorda@uni.lu}
\vspace{10pt}
\begin{indented}
\item[]April 2023
\end{indented}

\begin{abstract}
In many physical situations, there appears the problem of reaching a single target that is spatially distributed. Here we analyse how stochastic resetting, also spatially distributed, can be used to improve the search process when the target location is quenched, \emph{i.e.} it does not evolve in time. More specifically, we consider a model with minimal but sufficient ingredients that allows us to derive analytical results for the relevant physical quantities, such as the first passage time distribution. We focus on the minimisation of the mean first passage time and its fluctuations (standard deviation), which proves to be non-trivial. Our analysis shows that the no-disorder case is singular: for small disorder, the resetting rate distribution that minimises the mean first passage time leads to diverging fluctuations---which impinge on the practicality of this minimisation. Interestingly, this issue is healed by minimising the fluctuations: the associated resetting rate distribution gives first passage times that are very close to the optimal ones.
\end{abstract}

%
\vspace{2pc}

%
\submitto{\NJP}
%
%
%

\section{Introduction}

Stochastic resetting is an intriguing framework in stochastic dynamics because of its game-changing strategy in Brownian search processes, with applications in a wide range of contexts beyond physics, such as animal foraging, site-binding search of biomolecules or chemical reactions~\cite{Evans20jphysa}. In search processes, a usual goal is the minimisation of the mean first passage time (MFPT), which is the average time---over the different realisations of the dynamics---for finding the target. The simplest situation arises when one considers a particle undergoing Brownian motion in one dimension, with an immobile target at a certain position $x_T$ and a Poissonian reset process with rate $r$ to a certain point $x_r$. Resetting makes the MFPT finite and, in addition, there appears an optimal resetting rate $\widetilde{r}$---the MFPT diverges in both the limits of no resetting ($r=0$) and of infinitely frequent resetting ($r \to \infty$). 

General intermittent strategies, which combine searching reactive phases with faster relocating non-reactive phases, have been shown to be significantly useful in the context of search problems~\cite{Benichou_OptimalHidden_2005,Benichou_twodim_2006,Oshanin_intrandwalk_2007,Moreau_DisorderedMedium_2007,Oshanin_efficient_2009,Rojo_intermittent_2010,Chupeau_cover_2015}. These strategies have been applied in both microscopic and macroscopic contexts, both for single- and multiple-target problems, as discussed in the review paper \cite{Benichou_ReviewIntermittent_2011}. In particular, stochastic resetting can be thought of as a particular instance of these intermittent strategies, in which---in its simplest formulation---the non-reactive phase is instantaneous.

Typically, the target is considered to be fixed in space, \emph{i.e.} a single target remains at the same position for all  realisations of the search process. Still, the position of the target 
may vary between realisations, \emph{i.e.} there may be some degree of randomness in the search process that renders every instance thereof unique. We refer to this situation, in which the location of our goal is uncertain, as a search process with disorder---employing a terminology that is usual in physics. More specifically, we focus on problems for which the location of the target is fixed for each realisation of the stochastic process, which thus gives rise to a \textit{quenched disorder}. 
This quenched disordered scenario naturally emerges in many contexts, such as environmental phenomena associated with foraging~\cite{Viswanathan99nature,Marion05jtheobio,Bartumeus05ecology,boyer_modelling_2010,Viswanathan2011,berger-tal_recursive_2015,Pal20prr}. Therein, animals look for nutrient or resources throughout an environment that is heterogeneous; the targets are static (quenched) but spatially distributed. Another classical disordered scenario is that found for the location of minima of a complex energy landscape, which possibly represents the main challenge in disordered systems such as glasses~\cite{Cavagna09physrep,Berthier11rmp,Charbonneau14nature,Folena22physicaa,Ros23arxiv}; a problem that becomes highly non-trivial especially in high-dimensional contexts. 

In contrast with quenched disorder, annealed disorder---also called dynamical disorder~\cite{zwanzig_rate_1990}---can also be considered. In the foraging context, annealed disorder involves a dynamic target, which changes its position over  the same time scale of the searcher. Interestingly, stochastic resetting has been very recently introduced for intermittent targets, i.e. targets that fluctuate between a reactive and a non-reactive state~\cite{mercado-vasquez_first_2019,biswas_rate_2023}. This kind of intermittent targets are often found in cell biology, from the accessibility of DNA-binding sites to the opening and closing of ion channels, and they can be regarded as a particular case of annealed disorder.  In this work, we mainly address search processes with quenched (or static) disorder, \emph{i.e.} we consider that the position of the target follows a certain static, time-independent, spatial distribution $p_T(x_T)$. In the same context of foraging, this can be understood from the dynamics of searcher being much faster than that of the target, which is thus assumed to be immobile after the animal starts the search.

The assumption of disorder for the target location raises two questions in the field of stochastic resetting: what is the effect of stochastic resetting on such a disordered search problem? Is it possible to choose a clever, space-dependent reset rate $r(x)$ to optimise the search process, given the distribution of the target position $p_T(x_T)$? Note that a homogeneous resetting rate cannot be the optimal choice in general. For a fixed, delta-distributed target, with resetting position $x_r=0$, the optimal resetting rate is heterogeneous: $r(x<0) \to \infty$ and $r(x>0)=0$, since this choice always prevents particles from moving away from the target.

The above discussion entails that, physically, it is expected that any quenched heterogeneous distribution of targets $p_T(x_T)$, stemming from the heterogeneity of the searcher's environment, has an optimal heterogeneous resetting rate $r(x)$ associated. Despite the intuitive character of this statement, few works have been devoted to the analysis of stochastic resetting in heterogeneous environments. A first attempt involved the optimisation of the distribution of the resetting position in presence of a non-resetting window around the resetting point for exponentially distributed targets~\cite{Evans11jphysa}. Therein, a transition in the optimal distribution was found out. While a general framework to obtain the 
stationary distribution in presence of heterogeneous resetting has been developed~\cite{Roldan17pre},
the MFPT has been considered in works where the resetting takes place only when the particle's position exceeds some threshold value~\cite{Plata20pre,DeBruyne20prl}, a problem that naturally appears when modelling catastrophic events like power blackouts. In the context of systems with discrete states, a random walker on complex networks with node-dependent resetting rate has been considered as well~\cite{Ye22jstatmech}. Heterogeneity has also been investigated introducing spatially-dependent diffusion coefficients~\cite{Wang21pre,Lenzi22physicaa,Sandev22jphysa}, in a somewhat alternative direction with respect to the present paper. Rigorous results have been derived for the MFPT of a spatially-dependent resetting problem with a given distribution of the target position, assuming that explicit solutions for the MFPT equation are known~\cite{Pinsky20spa}. Heterogeneity thus emerges as a promising research ground in stochastic resetting and search processes.

The objective of this work is to investigate the effect of quenched disorder on the first passage time problem. Analytical results are typically hindered by the challenging mathematical structure of the optimisation problem. Therefore, we assume here a simple dichotomous time-independent distribution of the target position $p_T(x_T)$: the distance to the target is fixed and the probability of the target of being to the right (left) of the resetting point is $p$ ($1-p$). Consequently, we also assume a constant dichotomous resetting rate, piecewise homogeneous in the two real half-lines. These choices will allow us to answer analytically the following questions: what are the optimal resetting rates? how is the MFPT reduced by this assumption? what are the  fluctuations of the first passage time, especially when the disorder is small, i.e. in the $p\ll1$ (or $1-p\ll 1$) limit? what is the asymptotic behaviour of the optimal solution in the same limit and the long time tail of the first passage time distribution? The answer to these questions represents a first step towards the generalisation to more realistic target distributions and to heterogeneous resetting processes.

\section{Model}

We consider a one-dimensional freely diffusing overdamped Brownian particle submitted to stochastic resetting to a given point $x_r$, with heterogeneous resetting rate $r(x)$. Thus, the probability to observe the particle at position $x$ at time $t$ given that the initial ($t_0=0$) position was $x_0$, \emph{i.e.}~the propagator $P(x,t|x_0)$, follows the (forward) dynamic law
\begin{eqnarray}    
\partial_t P(x,t,|x_0)= & D \partial_x^2 P(x,t|x_0)- r(x) P(x,t|x_0) 
\nonumber \\ &
+ \delta (x-x_r) \int \!\de y \,r(y) P(y,t|x_0),
\label{eq:forward}
\end{eqnarray}
where $D$ is the diffusion constant. On the right hand side (rhs) of the evolution equation, each term has a neat physical meaning. 
Namely, the first term stands for free diffusion, the second one is a loss term due to resetting, and the last one is a gain term due to the particles that have reset from any point to $x=x_r$. To the purpose of the work, a piecewise constant resetting rate is assumed
\begin{equation}
\label{eq:r-piece}
r(x)= r_+ \Theta(x-x_r) +r_- \Theta(x_r-x) = 
\left\{ \begin{array}{@{\kern2.5pt}ll}
    \hfill r_+, & \mbox{if }  x>x_r,\\
    \hfill r_-, & \mbox{if }  x<x_r,
\end{array}
\right. 
\end{equation} 
where $\Theta$ stands for the Heaviside function. Our assumption of a fixed resetting location at $x=x_r$ is especially appealing---in addition to being the simplest---in the context of foraging problems, where the searching animal has a single home to come back.
Furthermore, the physically relevant information for our purposes is the 
distribution of the target location with respect to the resetting location; one can therefore safely fix the resetting location at the origin without any loss of generality.

Equation \eref{eq:forward} has to be complemented with initial and boundary conditions. The initial condition for the propagator is, of course, $P(x,0|x_0)=\delta(x-x_0)$. 
Boundary conditions depend on the physical constraints. On the one hand, in the absence of any constraint, the probability current vanishes at $x \to \pm\infty$, \emph{i.e.}~
\begin{equation}
\label{eq:free-bc}
\lim_{x \to \pm \infty} \partial_x P(x,t|x_0) = 0.
\end{equation}
In the long-time limit, the corresponding solution $P(x,t|x_0)$ 
tends to the non-equilibrium stationary state
\begin{eqnarray}  
\label{eq:sol_P_s}
    P_s(x) \equiv \frac{\alpha_+ \alpha_-}{\alpha_+ + \alpha_-} &\left[ e^{-\alpha_+ (x-x_r)} \Theta(x-x_r) 
    +  e^{\alpha_- (x-x_r)} \Theta(x_r-x) \right],
\end{eqnarray}
where the competition between diffusion and resetting defines the length scales $\alpha^{-1}_\pm = \sqrt{D/r_\pm}$. Note that  \eref{eq:sol_P_s} can be explicitly obtained due to the simple choice for $r(x)$ in  \eref{eq:r-piece}. An explicit solution  of equation \eref{eq:forward} is not available for arbitrary $r(x)$, which would define an associated space-dependent length scale $\alpha(x)^{-1}=\sqrt{D/r(x)}$.

On the other hand, if there is just one absorbing boundary at $x=x_T$, the boundary conditions to impose are
\begin{eqnarray}
P(x_T,t|x_0)=0 , \quad
\lim_{x \to {\scriptsize \textrm{sgn}}(x_r-x_T)\infty} \partial_x P(x,t|x_0) = 0 ,
\label{eq:abs-bc-inf}
\end{eqnarray}
standing $\textrm{sgn}(x)$ for the sign function. In our study, it suffices to impose the second condition in equation~\eref{eq:abs-bc-inf} and not necessarily $\lim_{x \to \pm \infty} \partial_x P(x,t|x_0) = 0$ because we will only consider situations where $x_0$ and $x_r$ are on the same side with respect to $x_T$. In a more general case, both limit conditions should be enforced. The case with just one absorbing boundary is the one addressed in this article, since we will look into the first passage problem to a target position with no further obstacles.   

Now we introduce the disorder, which is a fundamental and novel ingredient in our modelling. We assume that $x_T$ is static but not fixed; instead,  it is  distributed according to a certain distribution $p_T(x_T)$---i.e. the disorder is quenched.  Note that equation~\eref{eq:forward}, complemented with equation~\eref{eq:abs-bc-inf},
is still valid, but governing the evolution of the propagator $P(x,t|x_0;x_T)$, where it is important to make  the dependence on $x_T$ explicit. The propagator for the ensemble is achieved by integrating over $x_T$,
\begin{equation}
P(x,t|x_0)= \int \de x_T\, p_T(x_T) P(x,t|x_0;x_T).
\end{equation}      

As already commented in the introduction, we use the terminology quenched disorder throughout our work, which is quite usual in the context of statistical physics, in opposition to annealed disorder. The disorder is quenched in our model, since $x_T$ is ``frozen'' for each realisation of the dynamics. Instead, annealed disorder would involve a target $x_T$ evolving with a certain dynamics, even within a single realisation. In our context, a simple example of annealed disorder would be obtained if the target were relocated to a new random position extracted from $p_T(x_T)$ after each resetting event. Despite its apparent similarity, this problem with annealed disorder is not equivalent to the quenched disorder situation considered here. For the sake of simplicity, we use the simplest non-trivial quenched disorder 
\begin{equation}
p_T(x_T)=p \,\delta(x_T-x_r-X) + (1-p) \,\delta(x_T-x_r+X) ,
\label{eq:2dis}
\end{equation}
with $0<p<1$ and $X>0$. The distance from the resetting position to the target is fixed and equal to $X$, whereas there is a certain probability $p$ ($1-p$) for the target to be to the right (left) of $x_r$. The heterogeneity of $p_T$ motivates the asymmetric choice with respect to spatial inversion of $x-x_r$ for the resetting rate in \eref{eq:r-piece}.

It is handy to use dimensionless variables in order to identify the natural units of our problem. Specifically, we take $\bar{x}=(x-x_r)/X$ and  $\bar{t}=t/(X^2/D)$. Consequently, $\bar{P}(\bar{x},\bar{t}|\bar{x}_0;\pm 1)=X P(x,t|x_0;\pm X)$ and $\bar{\alpha}(\bar{x})=X \alpha(x)$. Then, equation~\eref{eq:forward} becomes
\begin{eqnarray}    
\partial_t P(x,t|x_0;x_T) =  &\partial_x^2 P(x,t|x_0;x_T) 
- \alpha(x)^2 P(x,t|x_0;x_T) \nonumber \\
&+ \delta (x) \int \!\de y \,\alpha(y)^2 P(y,t|x_0;x_T) ,
\label{eq:forward-adim}
\end{eqnarray}
in dimensionless form, where all bars have been removed in order to avoid cluttering our formulas. The boundary conditions in equation~\eref{eq:abs-bc-inf} %
become
\begin{eqnarray}
P(x_T,t|x_0;x_T)=0, \quad 
\lim_{x \to -\scriptsize \textrm{sgn} (x_T) \infty} \partial_x P(x,t|x_0;x_T) = 0,
\label{eq:abs-bc-adim}
\end{eqnarray}
where $x_T = \pm1 $ and
\begin{equation}
    \alpha(x)= \alpha_+ \Theta(x)+ \alpha_- \Theta(-x).
\end{equation}
Averaging over the disorder \eref{eq:2dis}, 
\begin{equation}
\label{eq:forward-av}
    P(x,t|x_0) = p\, P_+(x,t|x_0) + (1-p)\, P_-(x,t|x_0), 
\end{equation}
where we have conveniently introduced  the notation
\begin{equation}
\label{eq:def_Ppm}
    P_{+}(x,t|x_0) \equiv P(x,t|x_0;x_T = + 1) \ , \; \; P_{-}(x,t|x_0) \equiv P(x,t |x_0;x_T = - 1).
\end{equation} 
As described in the introduction, our focus is not the forward evolution but the first passage problem, which is formulated from the backward evolution in the next section. Nevertheless, the separation introduced in equation~\eref{eq:forward-av}, where the observable of interest is achieved from the average of the quenched contributions, each of them verifying its own equation, \emph{e.g.}~equation~\eref{eq:forward-adim} with boundary conditions~\eref{eq:abs-bc-adim} 
will be exploited along the rest of this work.

\section{First passage problem}
\label{sec:FPT}

From now on, we aim at solving the first passage problem to the disordered target. To accomplish this task, we start from the backward evolution equation for $P_{\pm}(x,t|x_0)$. Using an analogous notation to that for $P_{\pm}(x,t|x_0)$ in \eref{eq:def_Ppm}, we define the survival probabilities
\begin{equation}
\label{eq:def_surv}
    S_{\pm}(t|x_0)=\int \!\de x P_{\pm}(x,t|x_0),
\end{equation} 
\emph{i.e.}~$S_{+}(t|x_0)$ is the probability for having avoided at time $t$ the absorption at $x_T=+1$ starting from $x_0$---similarly for $S_-(t|x_0)$ with $x_T=-1$. The corresponding distributions of  first passage times to $x_T$ starting from $x_0$ are
\begin{equation}
\label{eq:def_FPTdist}
    f_{\pm}(t|x_0)=-\partial_t S_{\pm}(t|x_0).
\end{equation} 

All these functions, $P_{\pm}(x,t|x_0)$, $S_{\pm}(t|x_0)$, and $f_{\pm}(t|x_0)$, fulfill the same backward equation below. For the sake of concreteness, we write the backward equation just for the first passage time distribution $f_{\pm}(t|x_0)$, which  plays a central role in this research, 
\begin{equation}
\label{eq:FPT-dist}
\partial_t f_{\pm}(t|x_0) =  \partial_{x_0}^2 f_{\pm}(t|x_0) 
- \alpha(x_0)^2 f_{\pm}(t|x_0) 
+ \alpha(x_0)^2 f_{\pm}(t|0).
\end{equation}  
A brief derivation of the backward equation is presented in \ref{ap:backward}. The first passage time distribution has to be complemented with the boundary conditions
\begin{eqnarray}
f_{\pm}(t|\pm 1)=\delta(t), \quad 
\lim_{x_0 \to \mp \infty} \partial_{x_0} f_{\pm}(t|x_0) = 0,
\label{eq:bc-FPT}
\end{eqnarray}
and the initial condition $f_{\pm}(0|x_0)=0$, which is a consequence of $S_{\pm}(0|x_0)=1$ for any $x_0$. Moreover, since the particle unavoidably reaches the target at long times, we have the normalisation condition
\begin{equation}
    \int_0^{\infty} \!\de t  f_{\pm}(t|x_0) = \underbrace{S_{\pm}(0|x_0)}_{1} - \underbrace{\lim_{t \to \infty } S_{\pm}(t|x_0)}_0 =1.
\end{equation}

It is important to keep in mind that the first passage time distribution of the ensemble $f(t|x_0)$ is obtained averaging over the disorder,
\begin{equation}
\label{eq:FPTdis-ens}
    f(t|x_0)= p\, f_{+}(t|x_0)+ (1-p) \,f_{-}(t|x_0). 
\end{equation}
This is a property that is directly transferred  to all moments of the distribution, as well as any linear combination thereof. In what follows, we just focus on $f_+(t|x_0)$ due to the left-right symmetry of our system: $f_-(t|-x_0)$ is obtained by exchanging $\alpha_+ \longleftrightarrow \alpha_-$ in $f_+(t|x_0)$. This invariance stems from a mirror symmetry in our system: when simultaneously exchanging $x_T \longleftrightarrow -x_T$ and $r_+ \longleftrightarrow r_-$, the obtained system is just the mirror image of the original one with respect to the resetting point $x_r=0$. 

For solving the quenched first passage time distribution, we define the moment generating function (Laplace transform)
\begin{equation}
    \phi_{\pm}(s|x_0) = \int_0^{\infty} \de t\, e^{-st} f_{\pm}(t|x_0) ,
\end{equation}
from which all the moments of the distribution, $\tau^{(n)}$, are derived:
\begin{equation}
\label{eq:tau-def}
    \tau^{(n)}_{\pm}(x_0)= \int_0^{\infty} \de t \, t^n f_{\pm}(t|x_0) =(-1)^n \left. \partial_s^n \phi_{\pm}(s|x_0) \right|_{s=0}.
\end{equation}
Making use of the initial condition $f_{\pm}(0|x_0)=0$, equation~\eref{eq:FPT-dist} yields
\begin{equation}
\label{eq:GMF}
s \phi_{\pm}(s|x_0) =  \partial_{x_0}^2 \phi_{\pm}(s|x_0) 
- \alpha(x_0)^2 \phi_{\pm}(s|x_0) 
+ \alpha(x_0)^2 \phi_{\pm}(s|0),
\end{equation}
which, together with equation \eref{eq:tau-def}, gives the recursive equation for the moments
\begin{equation}
\label{eq:taus}
    -n \tau^{(n-1)}_{\pm} (x_0) =  \partial_{x_0}^2 \tau^{(n)}_{\pm}(x_0) 
    - \alpha(x_0)^2 \left[ \tau^{(n)}_{\pm}(x_0) - \tau^{(n)}_{\pm}(0) \right],
\end{equation}
with $\tau_{\pm}^{(0)}(x_0)=1$. 

Instead of solving equation~\eref{eq:taus} for the moments iteratively, we address the first passage time problem in a more compact way by solving equation~\eref{eq:GMF} for $\phi_+(s|x_0)$ at both sides of the resetting point.  Specifically, the boundary conditions along with continuity for the function and its derivative at $x_0=0$ must be imposed, as well as the consistency condition stemming from the local contribution at the resetting point---see \ref{ap:Lap-sol} for the detailed solution. Herein, for the sake of concreteness, we focus on the first passage problem starting from the resetting position, \emph{i.e.}~$x_0=0$. The result for the quenched moment generating function is
\begin{equation}
\label{eq:GMF-sol}
    \phi_+(s|0)= \left[  \frac{ \alpha_+^2 + s \cosh \left( \sqrt{s+\alpha_+^2} \right) }{s+\alpha_+^2} 
     +\frac{s \sinh \left( \sqrt{s+\alpha_+^2 }\right)}{\sqrt{s+\alpha_+^2}\sqrt{s+\alpha_-^2}} \right]^{-1}.
\end{equation}
The behaviour of the tails of the first passage time distribution is derived by carrying out an asymptotic analysis of the moment generating function. Specifically, one gets
\begin{equation}
\label{eq:asymp+1}
    f_+(t|0) \sim \left. \frac{(s+\alpha_+^2)(s+\alpha_-^2)}{\partial_s \zeta(s,\alpha_+,\alpha_-)} e^{st}\right|_{s=s^*_+}, \quad t\to \infty,
\end{equation}
where 
\begin{equation}
    \zeta(s,\alpha_+,\alpha_-) \equiv\frac{(s+\alpha^2_+)(s+\alpha^2_-)}{\phi_+(s|0)}  ,
\end{equation}
and $s^*_+$ is the pole with the highest real part of the moment generating function~\eref{eq:GMF-sol}, which is always real and negative. See~\ref{ap:uni-pole} for the detailed technical derivation. As already stated before, $\phi_{-}(s|0)$  is obtained from $\phi_{+}(s|0)$ by exchanging $\alpha_+ \leftrightarrow \alpha_-$. Therefore, we have that
\begin{equation}
\label{eq:asymp-1}
    f_-(t|0) \sim \left. 
        \frac{(s+\alpha_-^2)(s+\alpha_+^2)}{\partial_s \zeta(s,\alpha_-,\alpha_+)} e^{st}\right|_{s=s^*_-},
\end{equation}
with $s^*_-$ being the zero of the function $\zeta(s,\alpha_-,\alpha_+)$ with the highest real part. Hence, equations~\eref{eq:FPTdis-ens}, \eref{eq:asymp+1} and \eref{eq:asymp-1} entail that the first passage time distribution is a weighted sum of two exponential functions for long times. Such a prediction has been validated by simulating the stochastic process for different values of the system parameters, as shown below. 

Figure~\ref{fig:crossover} compares our theoretical predictions for the first passage time distribution with numerical results. In each panel, a different value of the disorder $p$ is displayed; the rest of parameters, \emph{i.e.}~$(\alpha_+,\alpha_-)$, are those minimising the MFPT for that value of $p$---derived later in section~\ref{sec:opt}. Simulation data (circles) correspond to $10^7$ trajectories for the parameters considered in each panel, integrated with a time step $\Delta t= 10^{-5}$ up to the first passage time; they show an excellent agreement with our theoretical predictions,  both asymptotic (solid green) and numerical Laplace inversion (dashed red). Although the asymptotic prediction is expected to work only in the limit of long times, it fails just at very short times. Therein, numerical inversion of the exact result still perfectly matches the simulations---as particularly shown in the inset of the second panel.

An especially interesting feature of figure~\ref{fig:crossover} is the apparent crossover between the two exponential decays corresponding to the long time behaviours of $f_{\pm}(t|0)$ in equations~\eref{eq:asymp+1} and \eref{eq:asymp-1}, provided that $p\ne \{0,1\}$. For $p=0$ (first panel), \emph{i.e.}~no disorder, only $f_{-}$ contributes and therefore the long time behaviour is exponential with $s^*_-$. For $0<p<0.5$ (rest of panels), the largest pole (smallest in absolute value) is $s^*_+$ for our choice of $(\alpha_+,\alpha_-)$; note that
$s^*_+$ vanishes in the limit $p \to 0$. This entails the crossover:  even for $p=0^+$, the long time behaviour is first dominated by $s^*_-$ (since the contribution from the $+$ mode is weighted with $p\ll 1$) but turns later to be  dominated by $s^*_+$. The crossover is not present for $p=1/2$, because $s^*_+=s^*_-$ for our choice of parameters. Note that each panel follows the same scale, this decision has been made to stress the emergence of a longer time exponential tail for non-zero disorder---on the first panel, it is clearly observed that the first passage distribution decays much faster than on the other panels. 
\begin{figure}
    \centering
    \includegraphics[width=\textwidth]{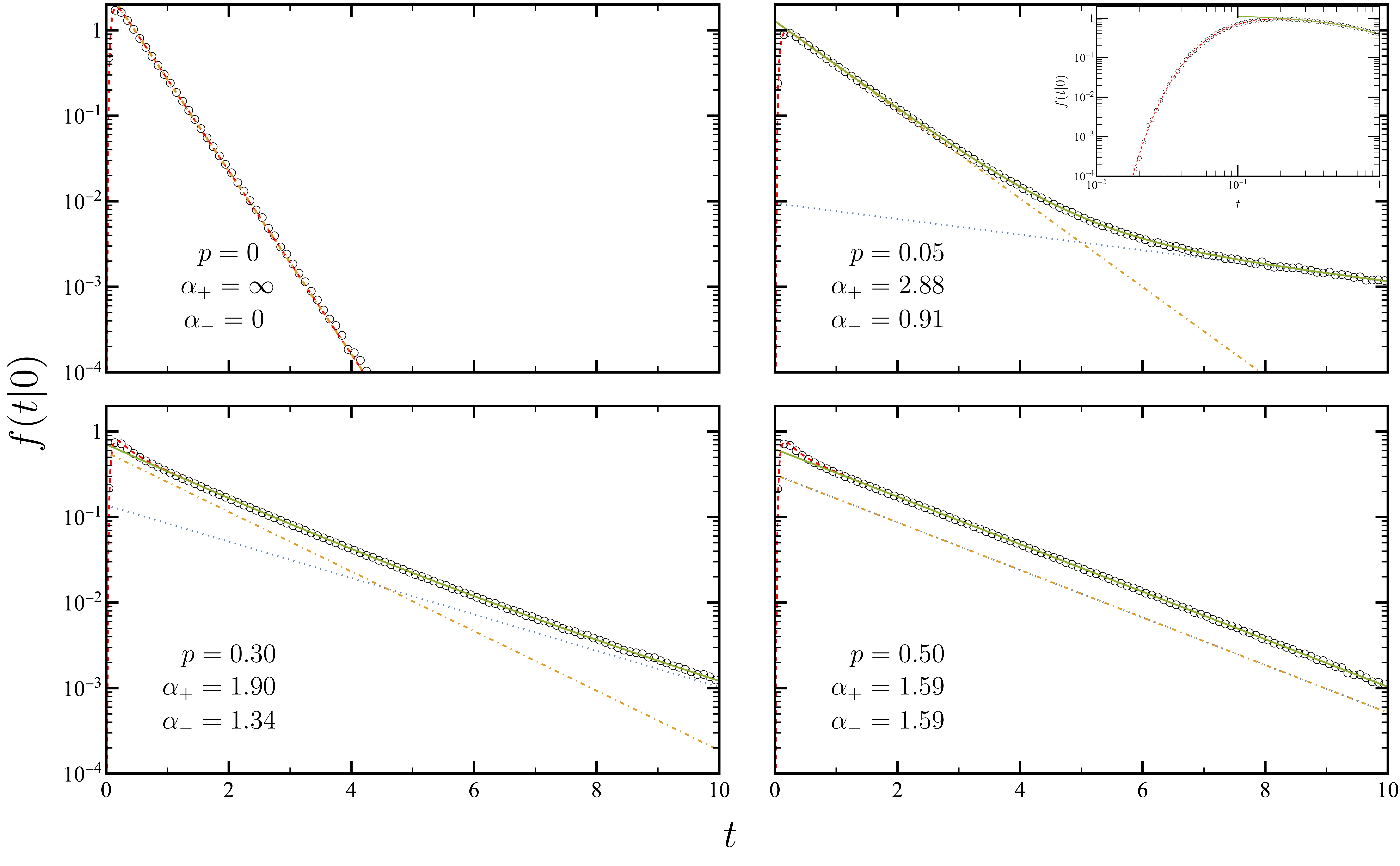}
    \caption{First passage time distribution for the disordered resetting system. 
    Specifically, we plot the results from the numerical simulation (circles), the numerical Laplace inversion of equation~\eref{eq:GMF-sol} (dashed red line), the asymptotic contribution from $f_+$ given by equation~\eref{eq:asymp+1} (dotted blue line),  the asymptotic contribution from $f_-$ given by equation~\eref{eq:asymp-1} (dotdashed orange line), and the sum of the two latter asymptotic contributions (solid green line). On the inset plot of the right top panel, we can check how the numerical Laplace inversion and simulations agree at short times and both lines overlap with the asymptotic estimation from $t=2\cdot 10^{-1}$ on.} 
    \label{fig:crossover}
\end{figure}

The mean and the second moment of the quenched contribution are obtained by using equations~\eref{eq:tau-def} and~\eref{eq:GMF-sol}---or, equivalently, solving equation~\eref{eq:taus} for $n=1,2$,
\begin{equation}
    \tau^{(1)}_+
    = \alpha^{-2}_+ \left(  \cosh \alpha_+ + \frac{\alpha_+}{\alpha_-} \sinh \alpha_+ -1 \right) \equiv F_1(\alpha_+ , \alpha_-),
    \label{eq:MFPT-sol}
\end{equation}
\begin{eqnarray}
        \tau^{(2)}_+  
        = \frac1{\alpha_-^3 \alpha_+^4} &\left\{ \left(\alpha_-^2-\alpha_+^2\right) \alpha_-+
   \left[\alpha_+^2-\alpha_-^2 \left(\alpha_-+3\right)\right] \alpha_+ \sinh \alpha_+ \right. \nonumber 
   \\ 
      & \quad  \left.- \left(\alpha_+^2+2 \alpha_--4 \alpha_+ \sinh \alpha_+\right) \alpha_-^2 \cosh \alpha_+ \right. \nonumber 
    \\ 
    & \quad \left.+ \left(\alpha_-^2+\alpha_+^2\right) \alpha_- \cosh \left( 2 \alpha_+ \right) 
    \right\} \equiv F_2(\alpha_+,\alpha_-).
\end{eqnarray}
Again, $\tau_{-}^{(n)}$ follows from $\tau_{+}^{(n)}$ by exchanging $\alpha_+ \longleftrightarrow \alpha_-$. Note that, in order to simplify the notation, we have omitted the argument of $\tau_{\pm}^{(n)}$, $\tau^{(n)}$; in this work, we only consider the case $x_0=x_r=0$. Averaging over the disorder, 
\begin{equation}
    \tau^{(n)}=p\,\tau^{(n)}_+
    +(1-p)\,\tau^{(n)}_- 
    .
\end{equation}

In the following, we mainly focus on the MFPT  and the standard deviation, $\tau^{(1)}$ and $\sigma_\tau$, respectively. The latter cannot be computed as an average between two contributions $\sigma_{\tau \pm}$ since it is not a linear combination of moments, we must use
\begin{equation}
\label{eq:STD-def}
    \sigma_\tau = \sqrt{\tau^{(2)} - \left( \tau^{(1)}\right)^2}.
\end{equation}
We have obtained exact results for $\tau^{(1)}$ and $\sigma_\tau$ as functions of $(p,\alpha_+,\alpha_-)$, but their explicit formulas have been deliberately obviated since they are not illuminating. For $\alpha_+=\alpha_-=\alpha_{\symm}$ (homogeneous resetting case),  the  expressions for the MFPT and the standard deviation are
\begin{equation}
    \tau^{(1)}_{\symm} = \frac{e^\alpha -1}{\alpha^2}, \quad \sigma_{\tau,\symm}= \frac{\sqrt{(2 \sinh \alpha - \alpha) e^\alpha}}{\alpha^2},
\end{equation}
which is a a robust result, valid regardless the value of $p$, due to the left-right symmetry. This result reasonably coincides with that obtained for homogeneous resetting in the absence of disorder~\cite{Evans11prl,Evans11jphysa}.

\begin{figure}
    \centering
    \includegraphics[width=0.49\textwidth]{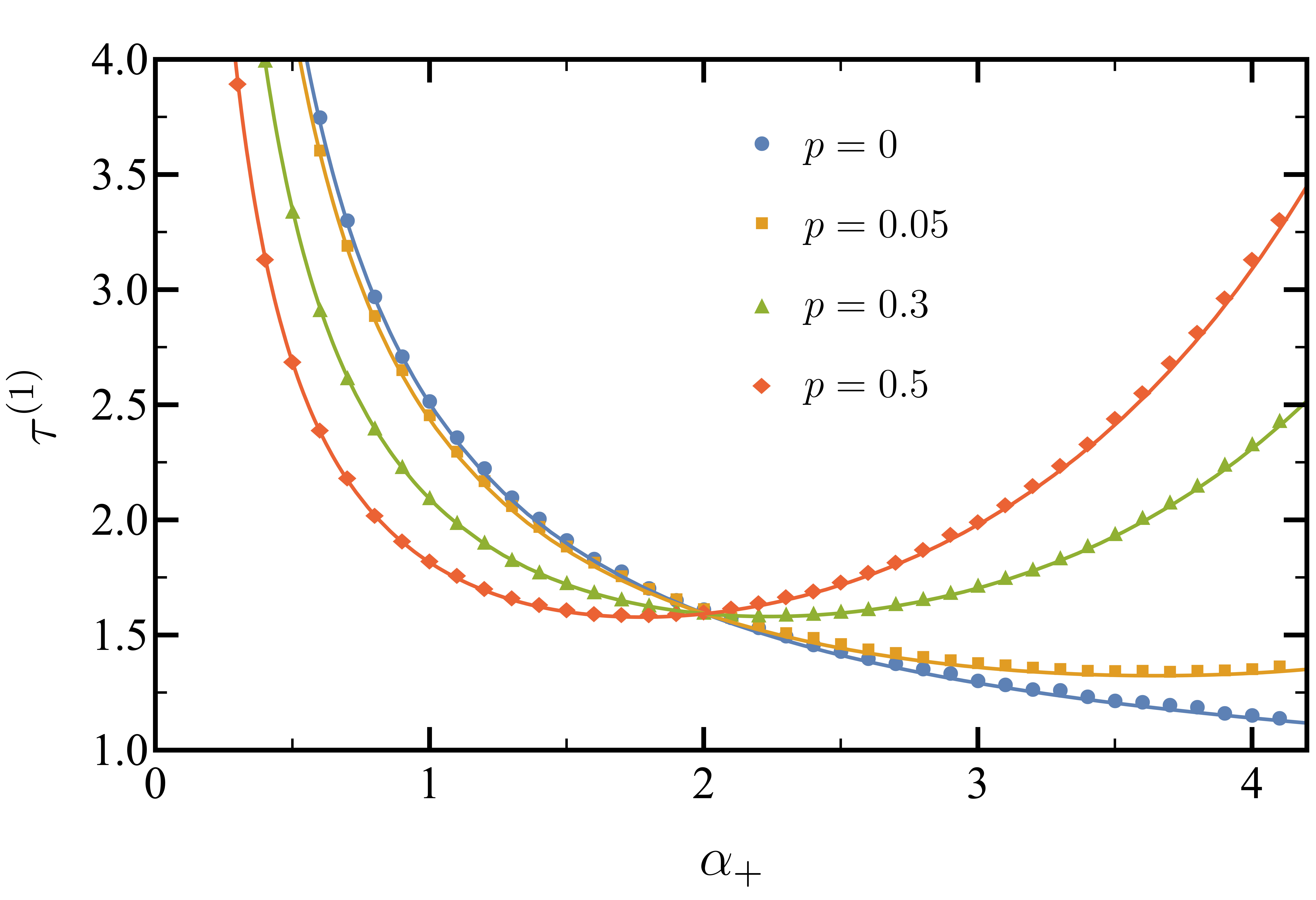}
    \includegraphics[width=0.49\textwidth]{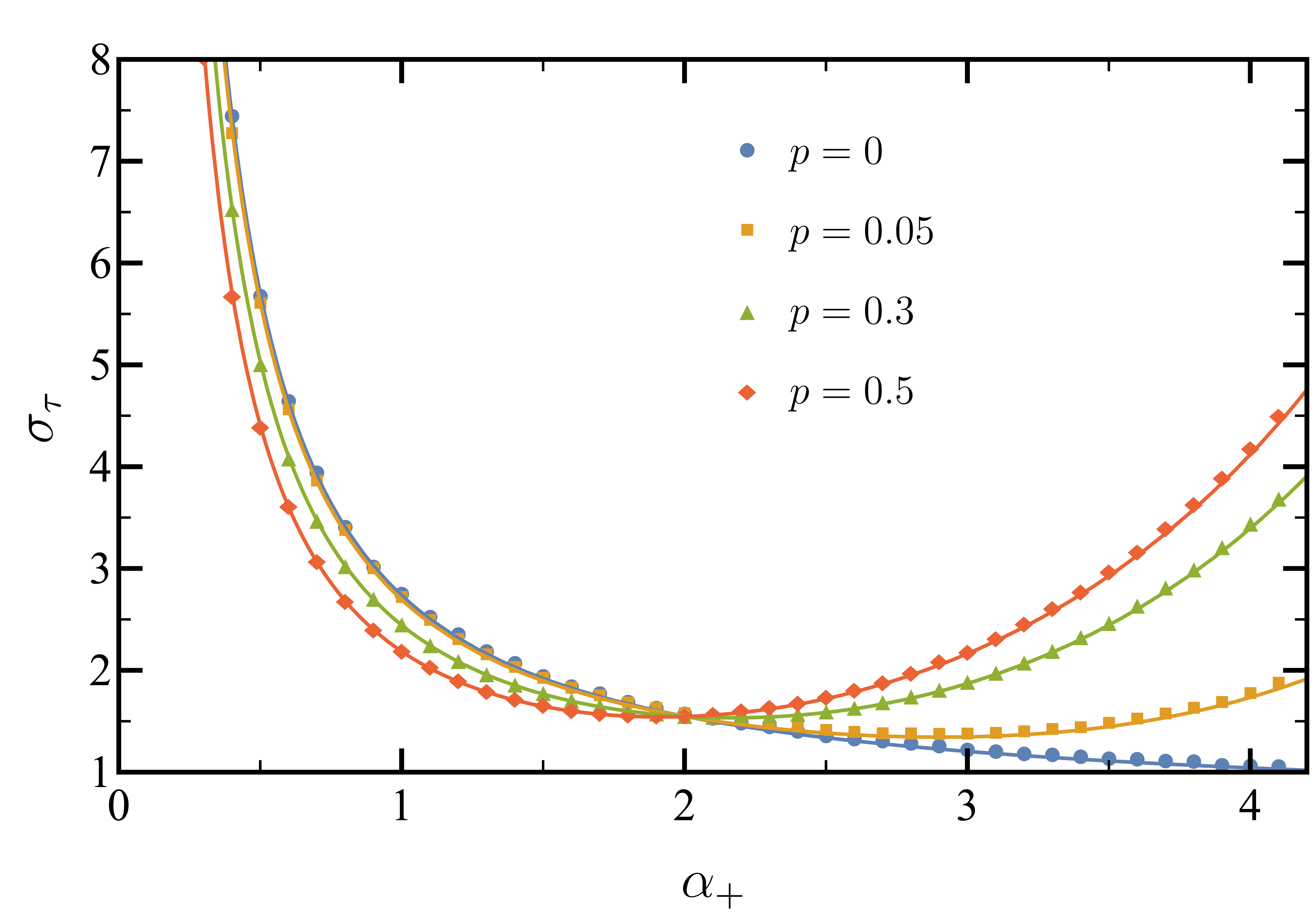}
    \caption{Mean first passage time (left panel) and its standard deviation (right panel) as functions of $\alpha_+$, at fixed $\alpha_-=2$. Simulation data (symbols) is compared with the exact analytical results (solid lines), showing a perfect agreement. Different values of disorder $p$ have been used to produce the different curves, as detailed in the legend.}
    \label{fig:mom-sim}
\end{figure}

In figure~\ref{fig:mom-sim}, we compare our theoretical predictions for $\tau^{(1)}$ and $\sigma_\tau$ with numerical simulations, again an excellent agreement is found. For the numerical results, a time step $\Delta t= 10^{-5}$ and $10^6$ trajectories have been used. Specifically, $\alpha_-$ has been fixed and the plots show the dependence of the observable of interest as a function of $\alpha_+$, for several values of $p$. All curves exhibit a minimum value as a function of $\alpha_+$, except those corresponding to $p=0$. Optimality of the observables  is discussed in section~\ref{sec:opt}.

\begin{figure}
    \centering
    \includegraphics[width=0.49\textwidth]{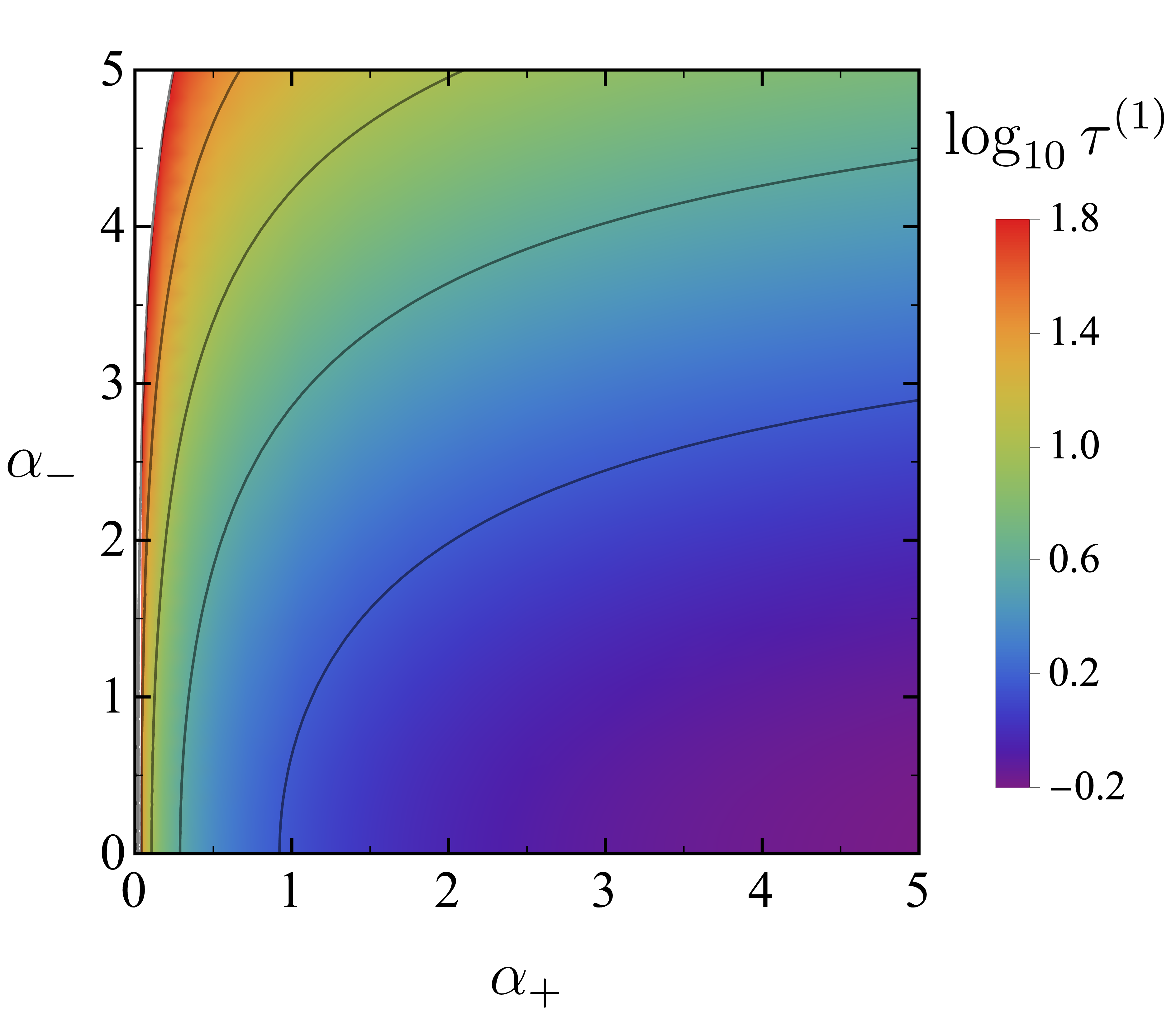}
    \includegraphics[width=0.49\textwidth]{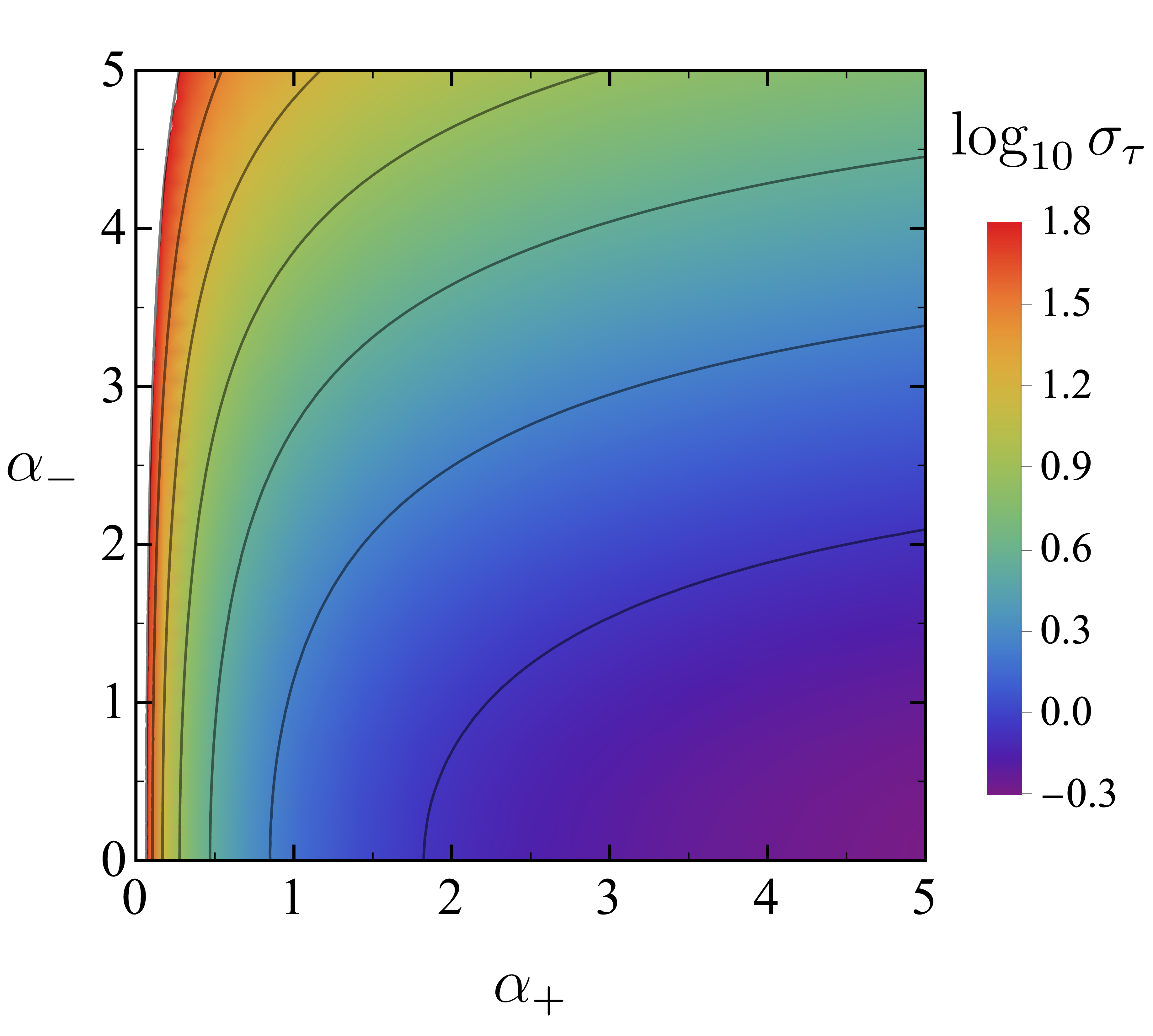}
    \includegraphics[width=0.49\textwidth]{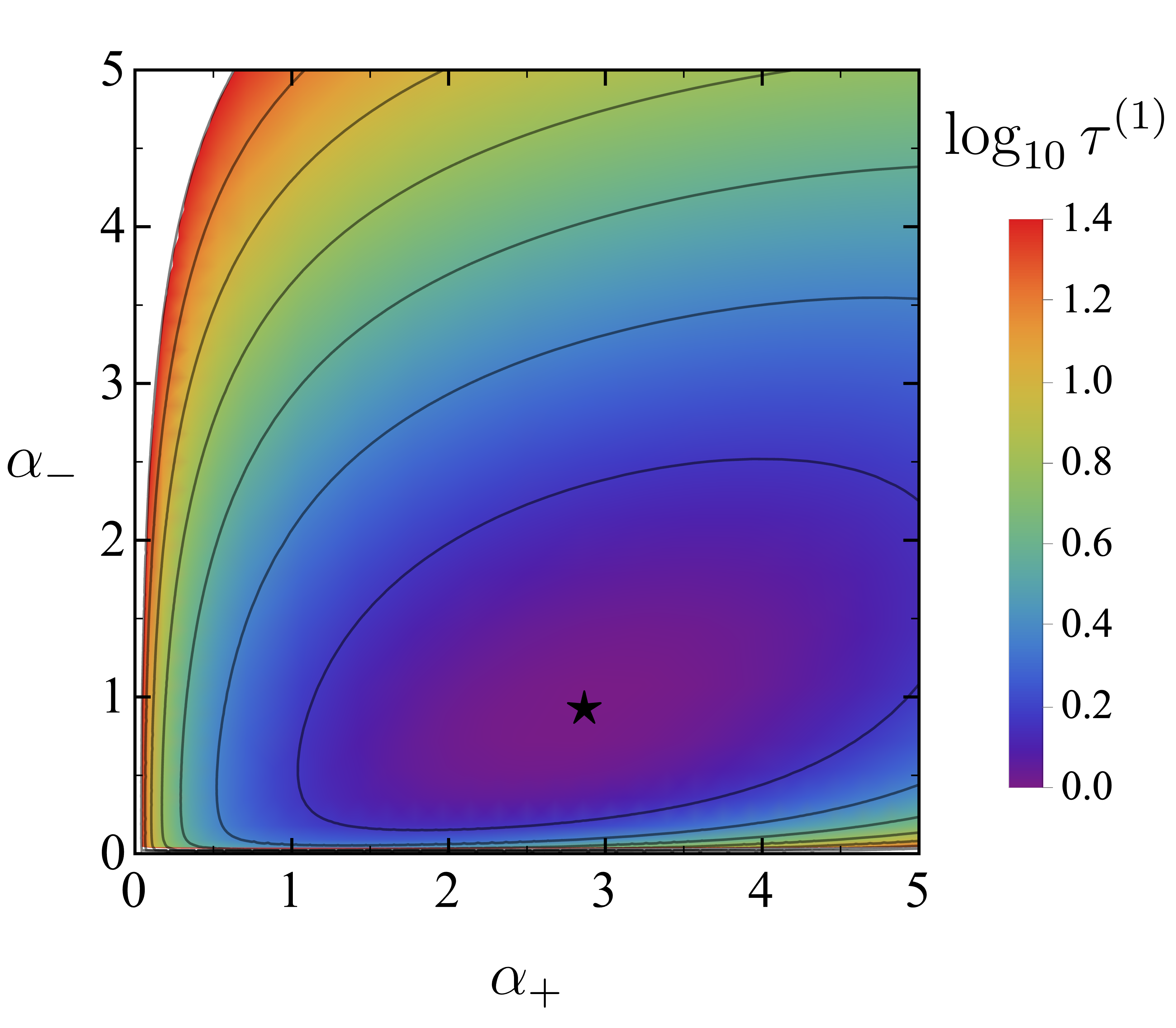}
    \includegraphics[width=0.49\textwidth]{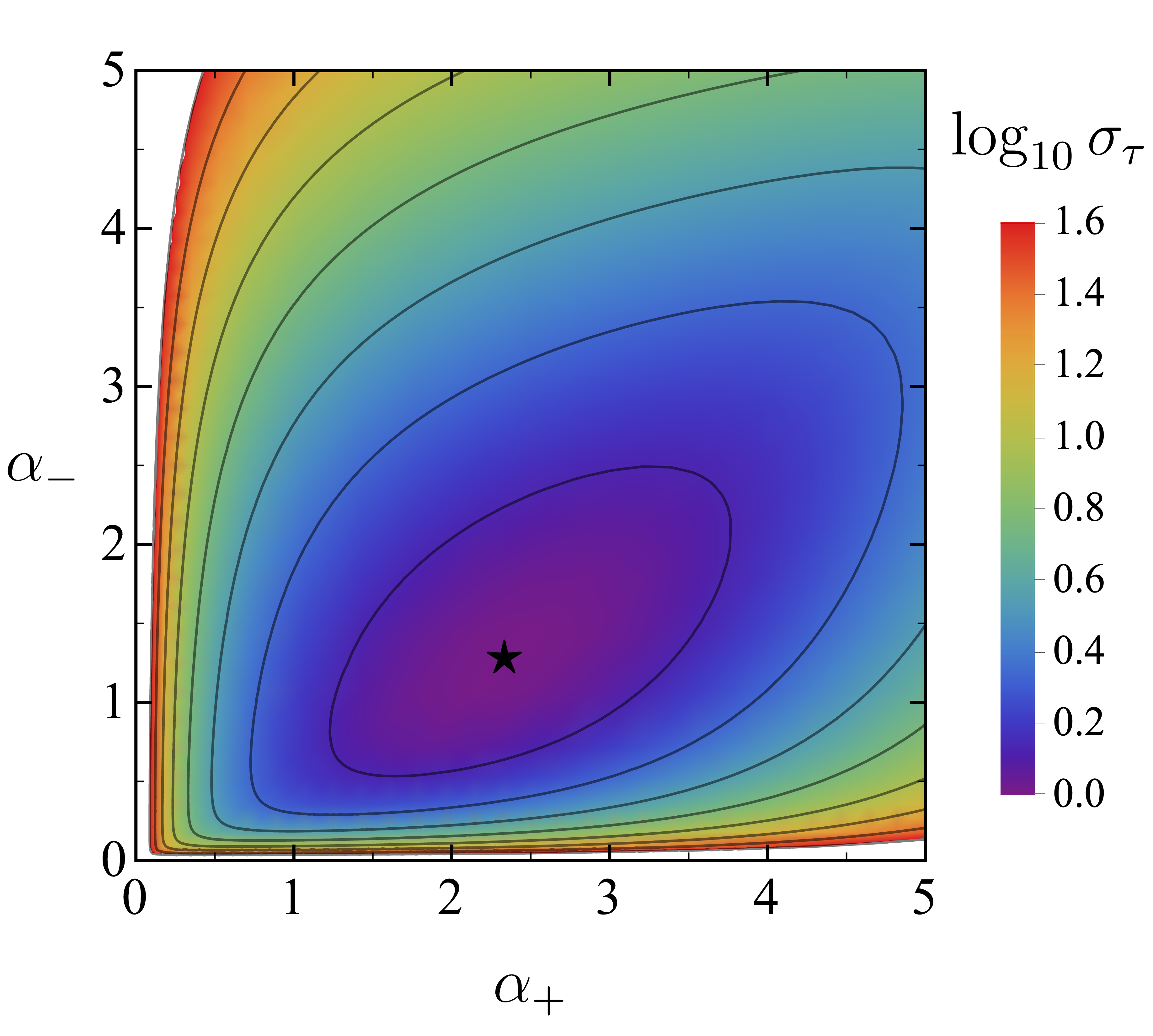}
    \includegraphics[width=0.49\textwidth]{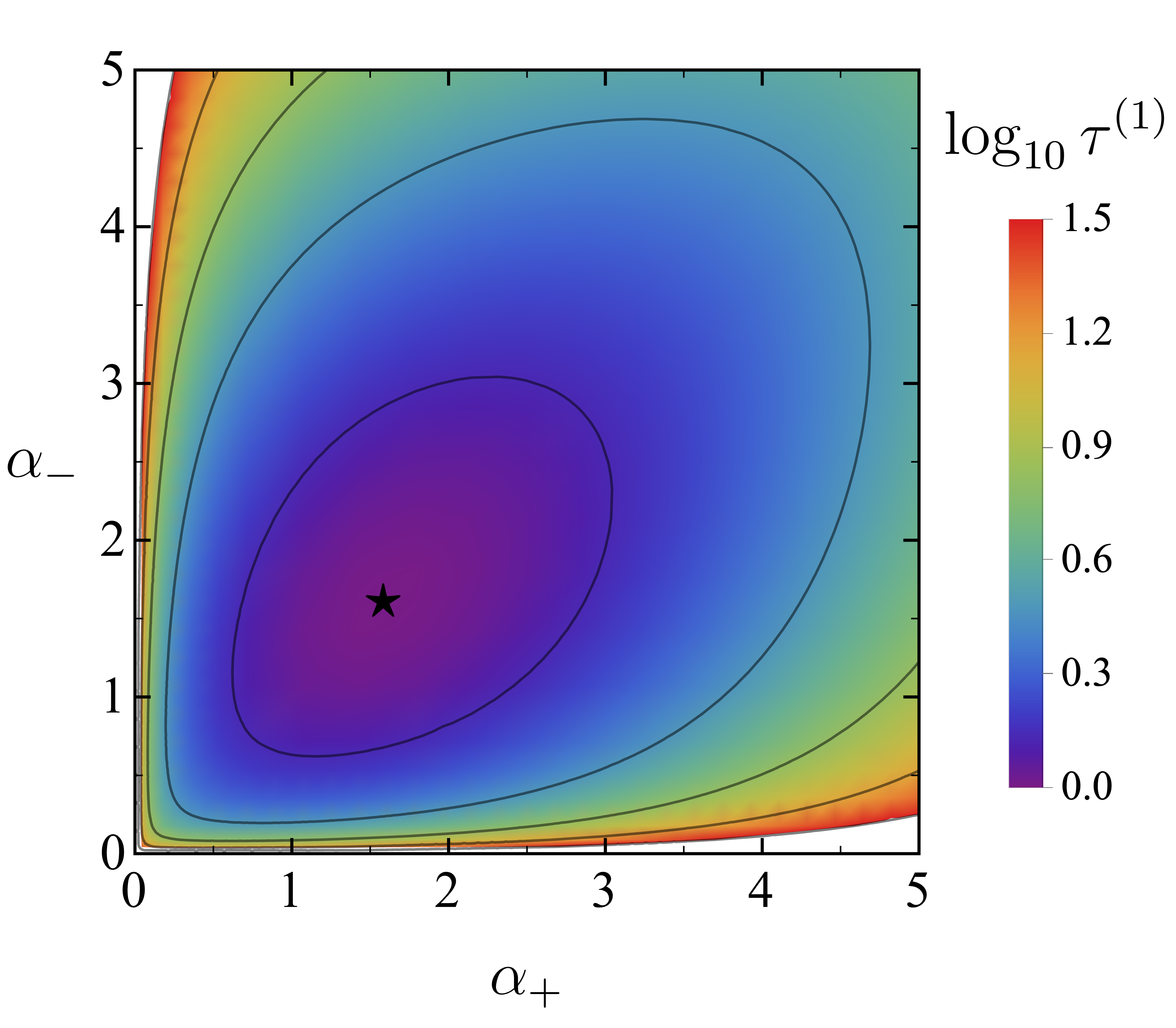}
    \includegraphics[width=0.49\textwidth]{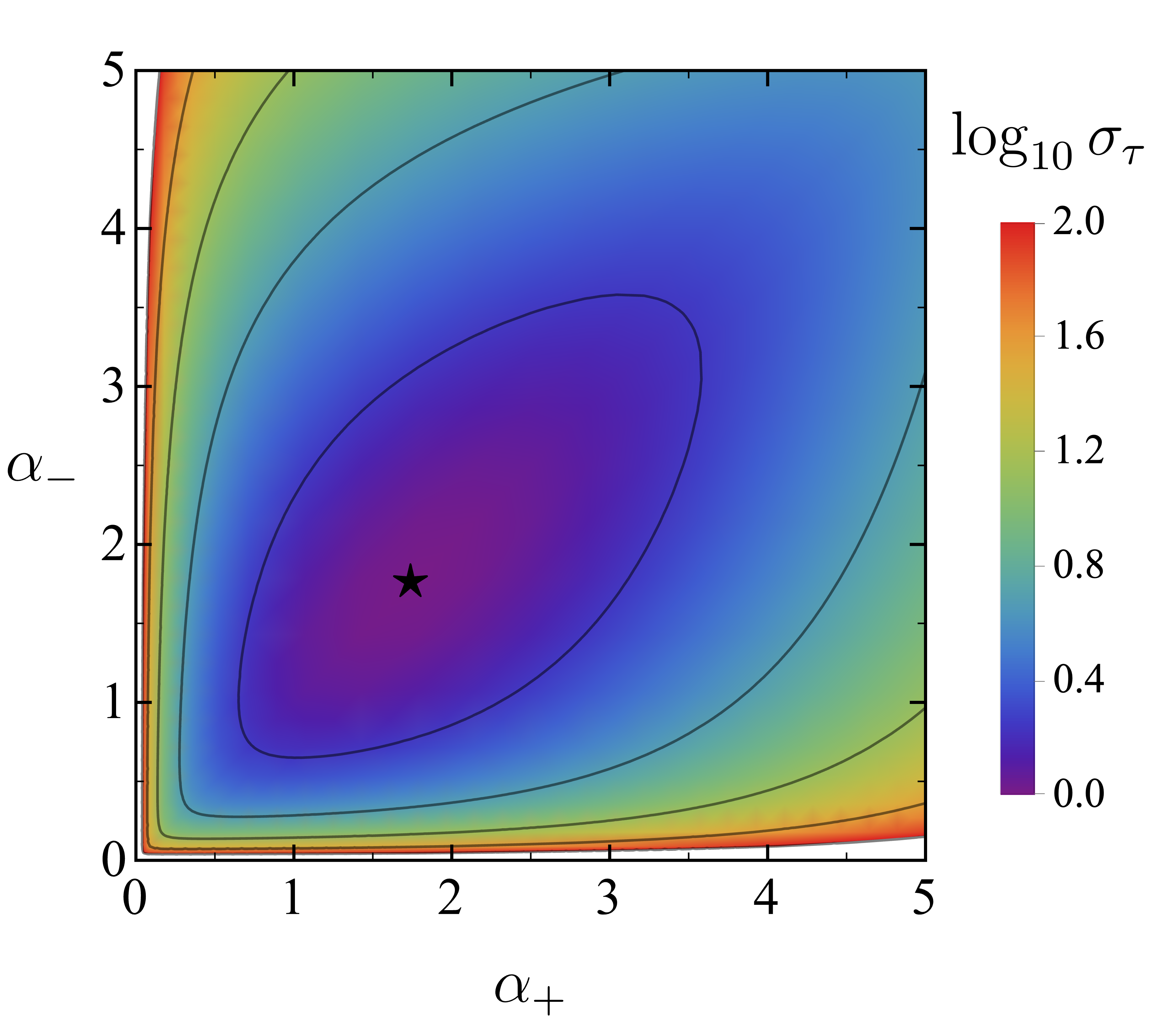}
    \caption{Density plots of the mean first passage time (left panels) and its standard deviation (right panels) on the $(\alpha_+,\alpha_-)$ plane.  They are obtained by the evaluation of our exact results in equations~\eref{eq:MFPT-sol}-\eref{eq:STD-def}. Different rows stand for different values of the disorder $p$, namely $p=\{0, 0.05,0.5\}$ from top to bottom. In each panel, white is used to indicate values larger than the maximum one in the corresponding colour legend, whereas the filled black star indicates the minimum---for $p=0$ it is reached at $\alpha_+\to \infty$ and $\alpha_-=0$.}
    \label{fig:mom-full}
\end{figure}

Figure \ref{fig:mom-full} provides a more complete picture of the behaviour of the MFPT and the standard deviation as functions of $(\alpha_+,\alpha_-)$, for different values of $p$. It is worth highlighting some interesting properties. As  expected, the panels for $p=0.5$ show reflection symmetry with respect to the line $\alpha_+=\alpha_-$. As a consequence of the properties of $F_{1,2}(\alpha_+,\alpha_-)$, for $p<1$, both the MFPT and the standard deviation diverge in the limits $\alpha_+ \to 0$ and/or $\alpha_- \to 0 $, except for the case $p=0$ where they remain finite for $\alpha_- \to 0$~\footnote{For $p=0$, the target is certainly to the left of $x_0=x_r$, and therefore it suffices to reset the trajectories straying to the right of $x_0$ to have a finite MFPT.}. Hence, excluding $p=0$, both objects develop non-trivial dependence on $(\alpha_{+},\alpha_-)$. 

\section{Optimisation of the physical quantities}
\label{sec:opt}

In this section,  the MFPT and its standard deviation are minimised. Moreover, the dependence of both the optimal values and their corresponding optimal resetting rates are looked into as a function of the disorder $p$.

\subsection{Optimal mean first passage time}

First, we focus on the MFPT. Mathematically, the problem reduces to finding the minimum of the function
\begin{equation}
\label{eq:MFPT1}
    \tau^{(1)}= p\, F_1(\alpha_+, \alpha_-) + (1-p) \,F_1(\alpha_-, \alpha_+),
\end{equation}
with $F_1(\alpha_+,\alpha_-)$ defined in Eq.~\eref{eq:MFPT-sol}. Then, it is needed to solve
\begin{equation}
\label{eq:MFPT-opt}
\left. \frac{\partial \tau^{(1)}}{\partial \alpha_+} \right|_{\raisebox{-0.8ex}{$\stackrel{\scriptstyle \alpha_+=\widetilde{\alpha}_+}{\scriptstyle \alpha_-=\widetilde{\alpha}_-}$}}=\left. \frac{\partial \tau^{(1)}}{\partial \alpha_-} \right|_{\raisebox{-0.8ex}{$\stackrel{\scriptstyle \alpha_+=\widetilde{\alpha}_+}{\scriptstyle \alpha_-=\widetilde{\alpha}_-}$}}=0
\end{equation}
for $( \widetilde{\alpha}_+,\widetilde{\alpha}_- )$. From now on, tilde notation refers to the solution of the minimisation problem of $\tau^{(1)}$. In the following, we study  both $(\widetilde{\alpha}_+,\widetilde{\alpha}_- )$ and its associated MFPT $\widetilde{\tau}^{(1)}$ as functions of $p$.

It is convenient to start with a physical intuition of the optimal problem. For $p=1$ ($p=0$), there is absolute certainty about the target position. Then, we expect the optimal resetting to forbid the exploration of the region on the opposite side of the target with $\alpha_- \to \infty$ $(\alpha_+ \to \infty)$, whereas the exploration of the correct direction is facilitated with $\alpha_+=0$ $(\alpha_-=0)$. This physical intuition is mathematically confirmed by taking into account that (i) $F_1$ is a monotonically increasing (decreasing) function with respect to its first (second) argument, and (ii) just one $F_1$ survives in equation~\eref{eq:MFPT1} in the extreme cases for $p$. This is clearly observed in the panel corresponding to $p=0$ in figure~\ref{fig:mom-full}, where the MFPT monotonically decreases (increases) with $\alpha_+$ ($\alpha_-$).

\begin{figure}
    \centering
    \includegraphics[width=0.7\textwidth]{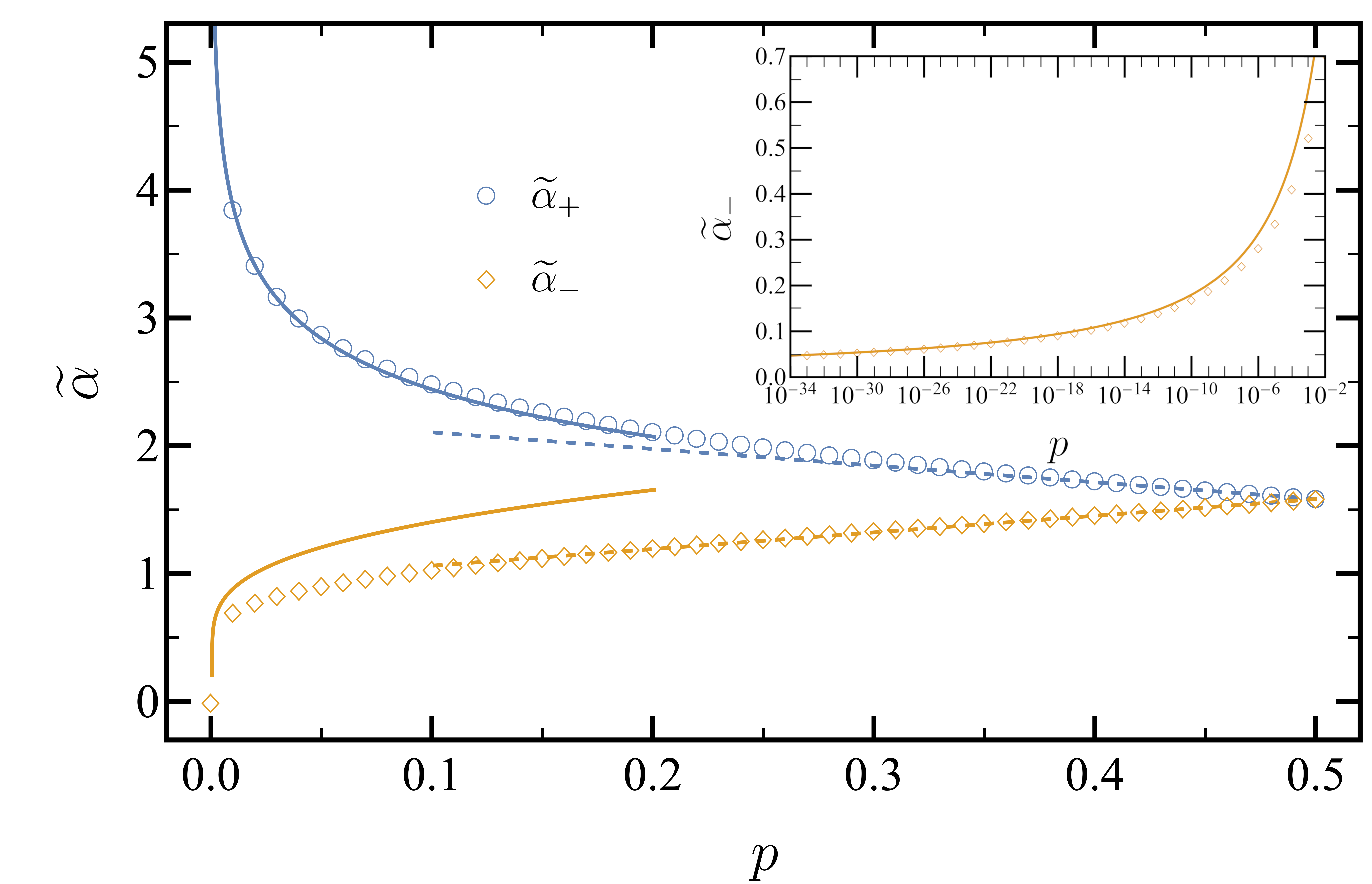}
    \includegraphics[width=0.7\textwidth]{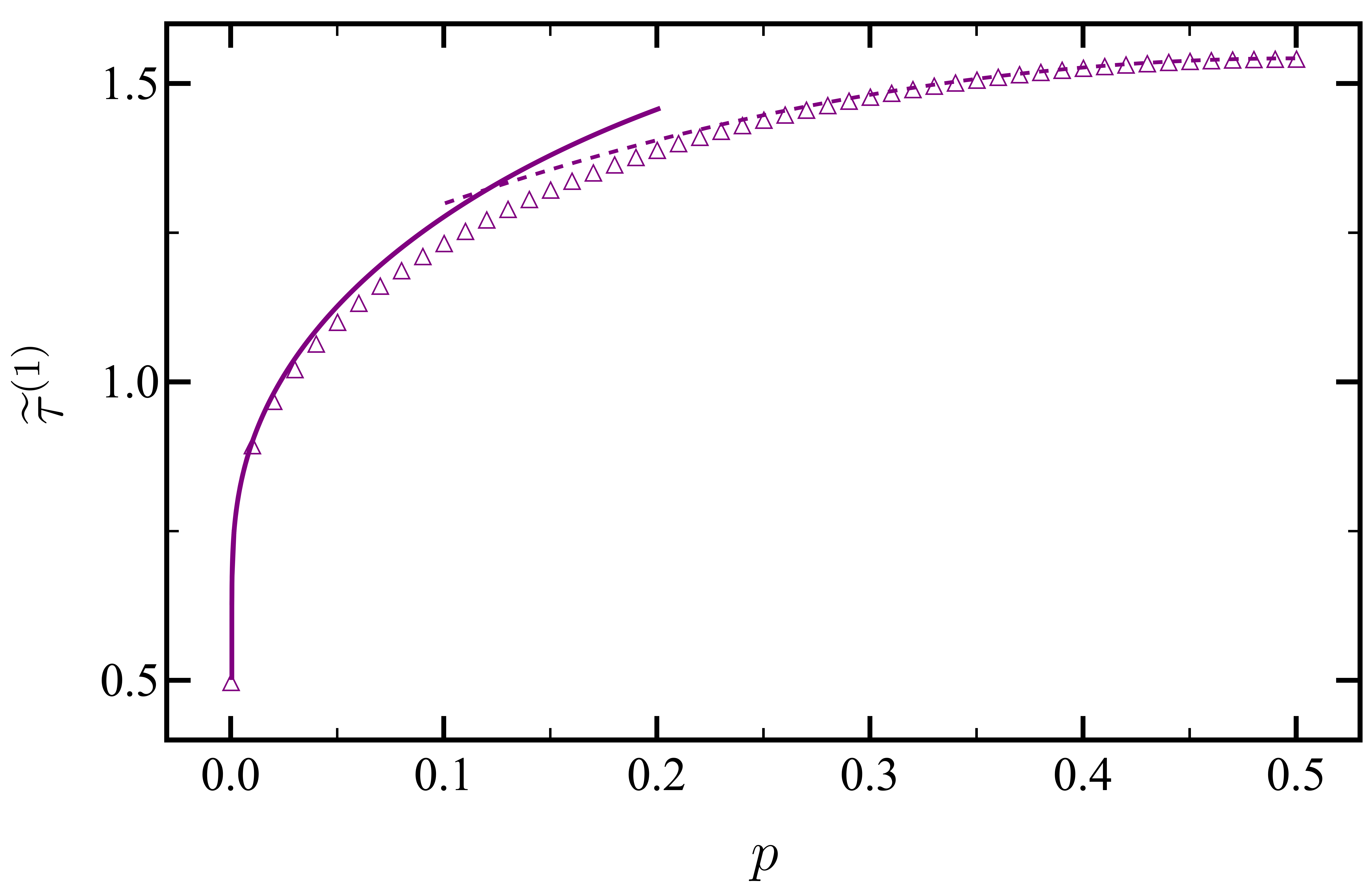}
    \caption{Minimum mean first passage time as a function of the disorder. Empty symbols stand for the numerical values obtained from minimising the MFPT, equation \eref{eq:MFPT-sol}, whereas the lines correspond to asymptotic expansions in the limits $p \to 0$ (solid line) and $p \to 1/2$ (dashed line). The top panel shows the optimal resetting rates $(\widetilde{\alpha}_+,\widetilde{\alpha}_-)$ as a function of $p$ ($\widetilde{\alpha}_+$: blue, $\widetilde{\alpha}_-$: orange). The inset is a zoom of the region close to $p=0$---note the logarithmic scale on the horizontal axis.  The bottom panel shows the associated MFPT $\widetilde{\tau}^{(1)}$, again as a function of $p$. The asymptotic expansions reproduce quite well the numerical solutions, the mathematical details thereof are provided in~\ref{ap:asymp-MFPT}.}
    \label{fig:minMFPT}
\end{figure}

For any $p \in [0,1] $, there is just one non-negative solution $( \widetilde{\alpha}_+,\widetilde{\alpha}_- )$ of equation~\eref{eq:MFPT-opt}. This solution has been numerically found  for any $p$, and also approximately through analytical expansions in the vicinity of $p=1/2$, $p=1$, and $p=0$. For the sake of clearness, the mathematical derivations as well as the specific expressions for the theoretical approximations are given in \ref{ap:asymp-MFPT}.  The optimal values of the resetting rates $( \widetilde{\alpha}_+,\widetilde{\alpha}_- )$ and the resulting minimum MFPT $\widetilde{\tau}^{(1)}$, along with their theoretical approximations, are shown in figure~\ref{fig:minMFPT} as functions of $p$. The accuracy of the analytical approximations are quite good, even for values of $p$ not so close to the ones used for the expansions. It is just in the limit as $p \to 0$ that the approximation for $\widetilde{\alpha}_-\to 0$ seems to fail, due to the quite rapid variation of $\widetilde{\alpha}_-$ in a narrow interval of $p$---this abrupt variation arises from the emergence of the disorder when changing from $p=0$ (no disorder) to $p\to0^+$ (weak disorder). Approximate analytical expressions are obtained for optimal resetting in~\ref{ap:asymp-MFPT}. The agreement between the approximation and the numerical results is reconciled when looking at the semi-logarithmic plot shown in the inset. The plots are restricted to the interval $p \in [0,1/2]$ because of left-right symmetry: for $p > 1/2$, the optimal rates are obtained by the exchange $\alpha_- \leftrightarrow \alpha_+ $ in those  for $1-p$, and the optimal MFPT $\widetilde{\tau}^{(1)}$ is thus symmetric with respect to $p=1/2$. 

\subsection{Optimal standard deviation of the first passage time}

Now we look into the minimisation of the standard deviation, \emph{i.e.}~look for the solution of the system of equations
\begin{equation}
\label{eq:STD-opt}
\left. \frac{\partial \sigma_\tau}{\partial \alpha_+} \right|_{\raisebox{-0.8ex}{$\stackrel{\scriptstyle \alpha_+=\widehat{\alpha}_+}{\scriptstyle \alpha_-=\widehat{\alpha}_-}$}}=\left. \frac{\partial \sigma_\tau}{\partial \alpha_-} \right|_{\raisebox{-0.8ex}{$\stackrel{\scriptstyle \alpha_+=\widehat{\alpha}_+}{\scriptstyle \alpha_-=\widehat{\alpha}_-}$}}=0,
\end{equation}
We use hat notation to refer to the optimisation of the standard deviation, in contrast to the tilde used for the minimisation of the MFPT. Since the analytical expression of $\sigma_\tau$ is quite cumbersome, analytical progress does not seem illuminating and we have addressed this minimisation problem numerically. 
\begin{figure}
    \centering
    \includegraphics[width=0.7\textwidth]{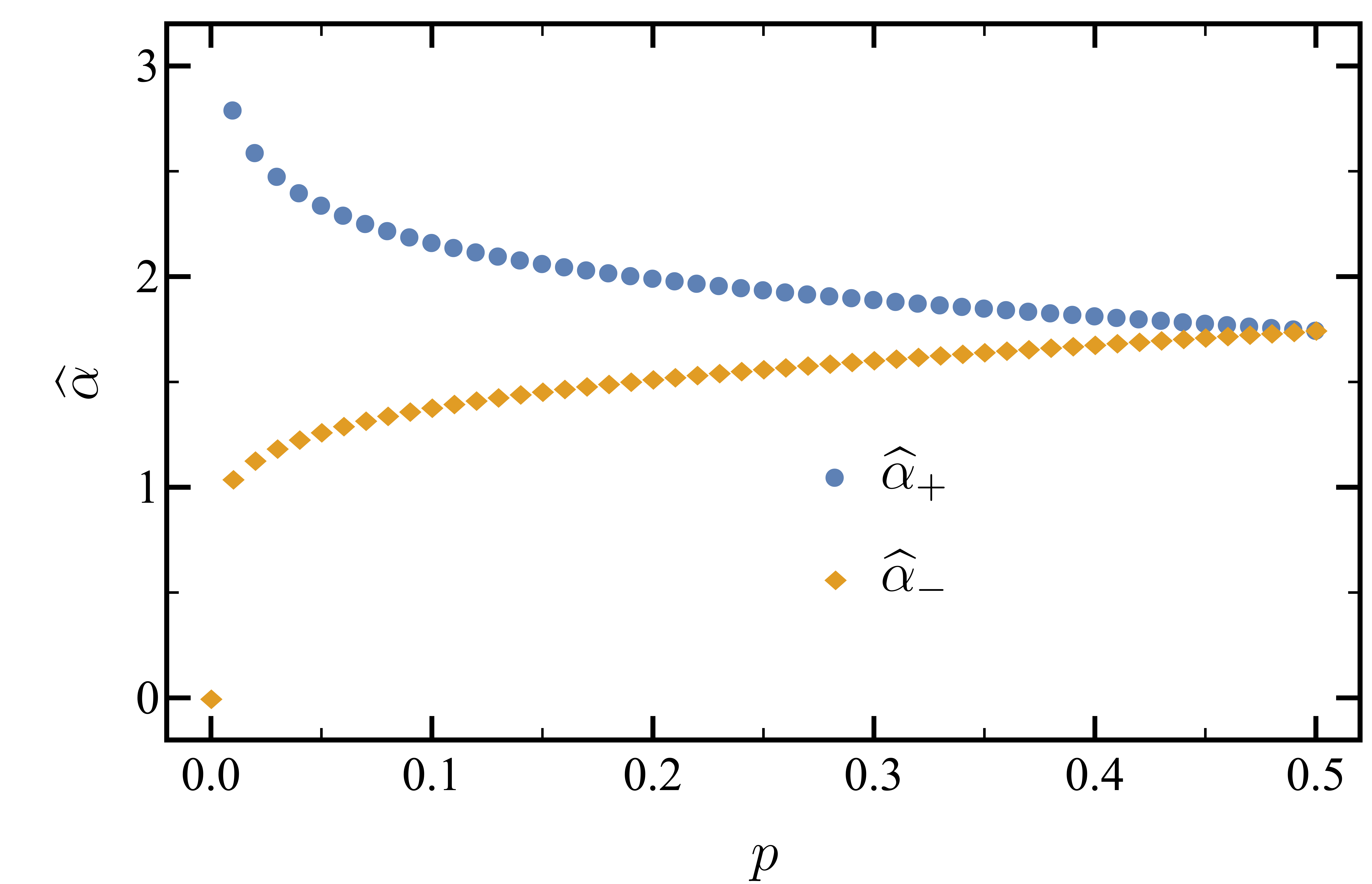}
    \includegraphics[width=0.7\textwidth]{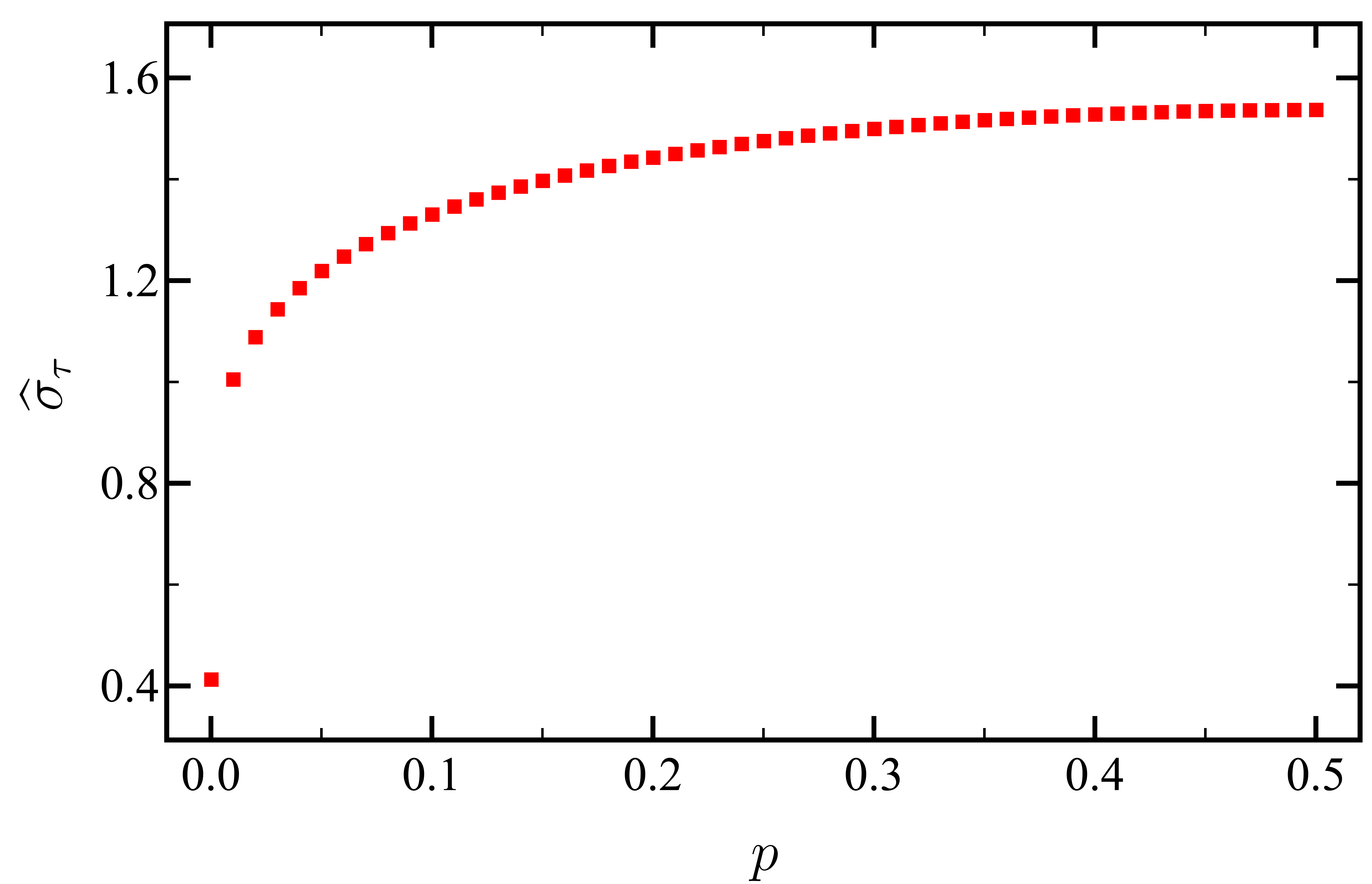}
    \caption{Minimum standard deviation of the first passage time for the disordered resetting system. Filled symbols stand for the numerical values obtained from minimising the standard deviation, as a function of the disorder $p$. The top panel shows the optimal resetting rates $(\widehat{\alpha}_+,\widehat{\alpha}_-)$ (filled blue circles and filled orange diamonds, respectively). The bottom panel displays its associated minimum value $\widehat{\sigma}_\tau$ (red rectangles), also as a function of $p$.}
    \label{fig:minSTD}
\end{figure}

In figure~\ref{fig:minSTD}, the optimal values of the resetting rates $(\widehat{\alpha}_+,\widehat{\alpha}_-)$ and the resulting minimum standard deviation $\widehat{\sigma}_\tau$ are shown as functions of $p$. As expected from direct inspection of the case $p=0$ in figure~\ref{fig:mom-full}, the asymptotic value for $\alpha_\pm$ obtained in the no-disorder limit $p=0$ is the same than that found in the minimisation of the MFPT.  However, the tendency towards the limit value are much slower here, as discussed in the following section. 

\subsection{Comparisons between the different optimisations}

\begin{figure}
    \centering
    \includegraphics[width=0.49\textwidth]{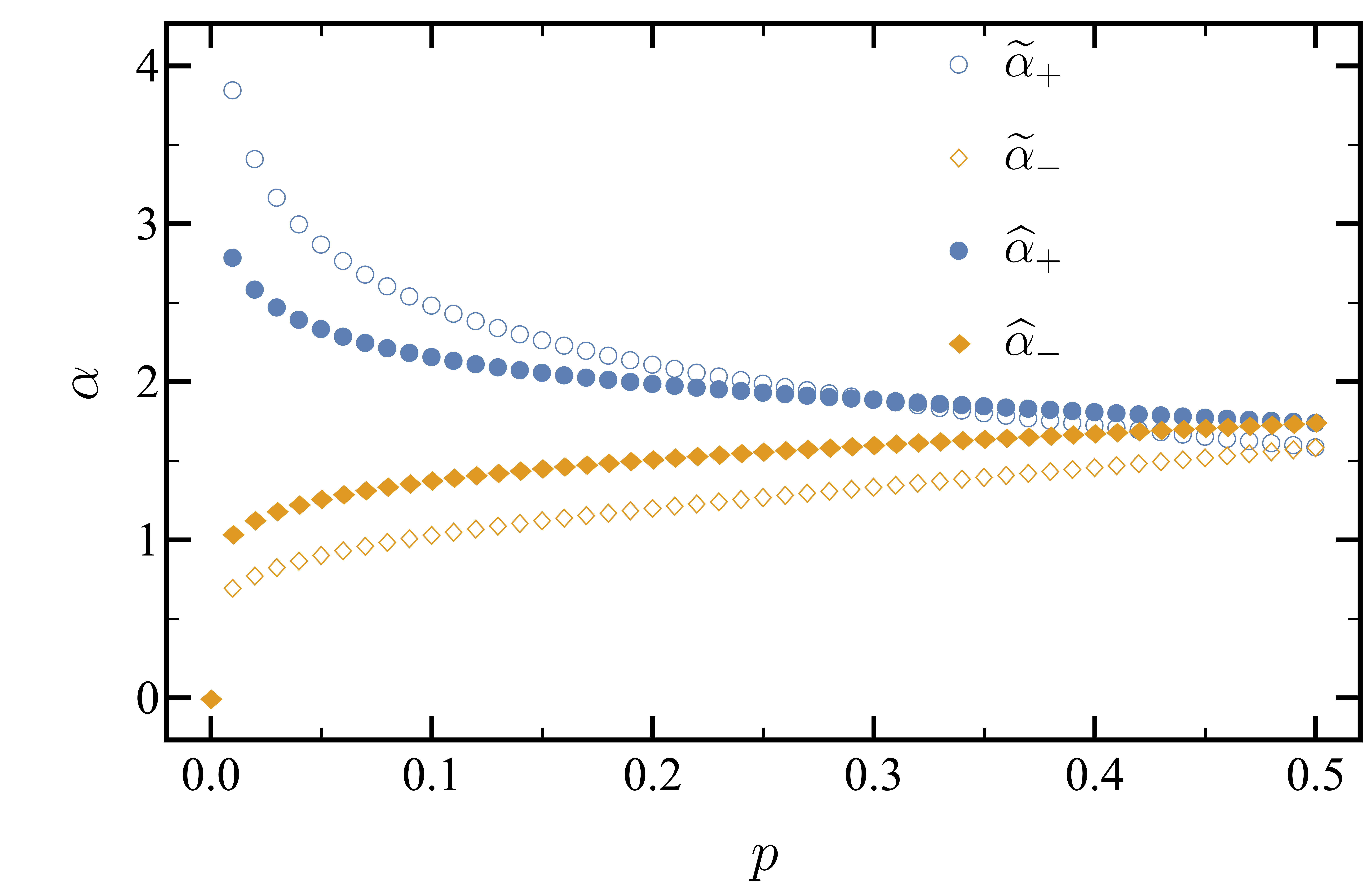}
    \includegraphics[width=0.49\textwidth]{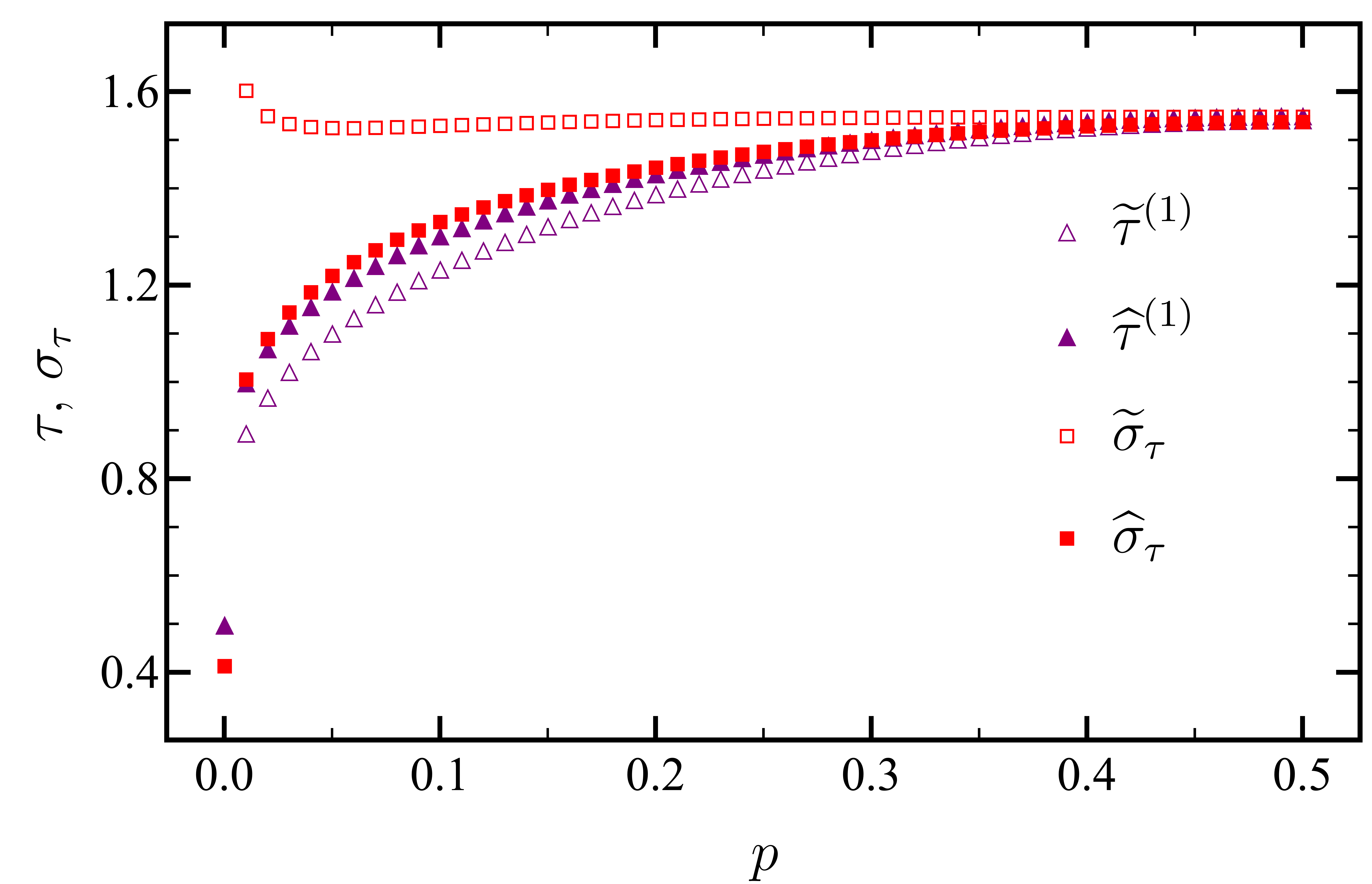}
    \caption{Comparison between the optimal solutions for the mean first passage time and the standard deviation, all of them obtained from solving numerically each quantity. On the left panel, we show the optimal resetting rates $(\alpha_+,\alpha_-)$ (blue circles and orange diamonds, respectively) when either MFTP or the standard deviation are minimised. On the other hand, on the right panel, we compare all the possible cases about the MFTP and standard deviation (purple triangles and red rectangles, respectively) when one of them is optimal. When the marker is filled indicates that the standard deviation is being minimised, whereas it is empty the MFTP is optimal.
    }
    \label{fig:comp}
\end{figure}

Now we compare the two optimisations we have just analysed, that of the MFPT and that of the standard deviation. In the left panel of figure~\ref{fig:comp}, we present together the optimal resetting rates $(\widetilde{\alpha}_+,\widetilde{\alpha}_-)$ and $(\widehat{\alpha}_+,\widehat{\alpha}_-)$ that minimise $\tau^{(1)}$ and $\sigma_\tau$, respectively. From a qualitative point of view, their behaviour seems to be quite similar, but the optimal resetting rates at minimum standard deviation $(\widehat{\alpha}_+,\widehat{\alpha}_-)$ have a slower dependence on $p$ than that of  the optimal resetting rates at minimum MFPT $(\widetilde{\alpha}_+,\widetilde{\alpha}_-)$. In the right panel of figure~\ref{fig:comp}, we compare the minimum MFPT $\widetilde{\tau}^{(1)}$ with the MFPT $\widehat{\tau}^{(1)}$ calculated at the rates that minimise the standard deviation; and also the minimum standard deviation $\widehat{\sigma}_\tau$ with the standard deviation $\widetilde{\sigma}_\tau$ calculated at the rates that minimise the MFPT. The standard deviation at minimum MFPT $\widetilde{\sigma}_\tau$ explodes in the vicinity of $p=0$, $\lim_{p \to 0^+} \widetilde{\sigma}_\tau= \infty$ despite $\widetilde{\sigma}_\tau$ is finite for $p=0$. 
This divergence stems from the contribution to the second moment of the less probable target---when evaluated at the extreme values of the resetting rates. The different behaviour of $\widehat{\alpha}_\pm(p)$ regularises the standard deviation $\widehat{\sigma}_\tau$; although the contribution of the second moment stemming from the less probable target still diverges, it remains finite when multiplied by $p$. The divergence of the standard deviation at finite MFPT stems from the interplay between the low value of $p$, which entails that trajectories with $x_T=1$ are very rare, and the divergent waiting times over such rare trajectories---as clarified in~\ref{ap:asymp-MFPT}.

It is interesting to note that the MFPT at minimum standard deviation $\widehat{\tau}^{(1)}$ in the right panel of figure~\ref{fig:comp} is not far from the minimum MFPT $\widetilde{\tau}^{(1)}$. This is remarkable: if one is interested in avoiding large fluctuations, one should employ the hat parameters minimising the standard deviation; in this way the MFPT is still very close to the optimal one.\footnote{Also, the minimisation of a combined figure of merit, like the sum of the MFPT plus the standard deviation, would be possible~\cite{Solon18PRL}.}
.

\section{Conclusions}
In this work, we have investigated the impact of a distributed target position in a minimal model of a stochastically resetting system. This has allowed us to model the uncertainty about the location of the goal of a searcher---in other words, the heterogeneity of its environment, and shed light on looking for realistic optimal search strategies. 

In our dichotomous one-dimensional model, the target has probability $p$ ($1-p$) of being to the right (left) of the resetting, which is also the initial, position of the searcher $x_r$. Likewise, the resetting rate is different to the right and to the left of $x_r$. 
The target position represents then a minimal realisation of a quenched disorder, \emph{i.e.} a fluctuating variable distributed among the realisations but static during a single trajectory.
Both exact analytical results and approximate asymptotic expressions in some limits have been obtained for this novel scenario, our theoretical approach has been validated by numerical results. The ``deterministic'' case $p=\{0,1\}$ has been proven to be singular when compared with the small randomness limit, \emph{i.e.}~$p \to \{0^+,1^-\}$.

With respect to the first passage time distribution, there appears a crossover between two different exponential behaviours, as a consequence of disorder. In the explored range of parameters, chosen by minimising the MFPT, this is especially significant for small non-vanishing disorder. Therein, a quite long exponential tail, stemming from the contribution of the less probable target, emerges. 

The minimisation of the MFPT and the standard deviation have revealed interesting features of optimal search strategies.
Of course, the optimal resetting rates with no disorder are those physically expected: the optimal resetting is maximal on the side opposite to the target, whereas there should be no resetting on the target side. Nonetheless, the behaviour at small $p$ is very different from the deterministic case: the standard deviation associated to minimum MFPT diverges. Thus, in a given trajectory, one may experiment very long first passage times at the resetting rates that minimise the MFPT, due to the large fluctuations, which impinges on the usefulness of this minimisation. Interestingly, the minimisation of the fluctuations, \emph{i.e.} of the standard deviation, gives first passage times very close to the minimum ones for all values of $p$. This demonstrates the importance of taking an adequate figure of merit, from a practical point of view it is better to minimise the standard deviation because very long first passage times become more rare. Checking if this fact still holds for more complex target distributions is an interesting prospect for further research.

Not only has this study provided new insights on optimal search strategies within resetting systems but also opened promising perspectives for the future of the field. We have focused on providing a minimal model, for which the novel first passage properties stemming from the disorder could be obtained in an exact way. However, there are many aspects still to be explored, as a consequence of the distributed target position: (i) the latter can be interpreted as a quenched disorder, and a search process in a quenched vs annealed case can be investigated in the framework of disordered systems~\cite{Montanari02prl,Castellani05jstatmech,Berthier11rmp}; (ii) the problem of finding an optimal resetting rate $\alpha(x)$ for a given target distribution $p_T(x_T)$ in $1d$ processes is still open~\cite{Pinsky20spa,Evans20jphysa}, and our work represents a first step in this direction; (iii) our results may be generalised to higher-dimensional systems, where we conjecture that the MFPT performance may increase significantly~\cite{Evans14jphysa}; 
(iv) prior information about the target distribution leads to more efficient search strategies; in the context of stochastic thermodynamics~\cite{Seifert12rpp,Parrondo15natphys,Peliti2021}, it would be interesting to address the role of fluctuation relations and the entropy of information in resetting systems~\cite{Fuchs16epl,Gupta20prl,Gupta22njp}---in analogy with the situation found in feedback controlled systems~\cite{Abreu12prl,barato_unifying_2014,ruiz-pino_information_2023};
(v) it is tempting to associate this information to an active particle exploring the environment with a persistent self-propulsion~\cite{Bechinger16rmp,Fodor18physicaa}, following recent approaches in resetting dynamics~\cite{Evans18jphysa,Tucci22pre}.

\section*{Data availability statement}

The data that support the findings of this study are openly available at the following URL/DOI: 
\href{https://github.com/fine-group-us/disordered-resetting}{https://github.com/fine-group-us/disordered-resetting}

\section*{Acknowledgements}

C.~A.~Plata acknowledges the funding received from European Union's Horizon Europe--Marie Sk\l{}odowska-Curie 2021 programme through the Postdoctoral Fellowship with Ref.~101065902 (ORION). G.~García-Valladares, C.~A.~Plata and A.~Prados acknowledge financial support from Grant PID2021-122588NB-I00 funded by MCIN/AEI/10.13039/501100011033/ and by ``ERDF A way of making Europe'', and also from Grant ProyExcel\_00796 funded by Junta de Andalucía's PAIDI 2020 programme. A.~Manacorda acknowledges the funding received from European Union's Horizon Europe--Marie Sk\l{}odowska-Curie 2021 programme through the Postdoctoral Fellowship with Ref.~101056825 (NewGenActive). 

\appendix

\section{Derivation of the backward equation for the first passage time distribution}
\label{ap:backward}

In search problems, where an agent aims at reaching a certain target, it is handy to use the backward approach. Through this formalism, the evolution of the probability density function is also analysed, but in terms of the initial values of the variables---instead of the values of the variables at time $t$ employed in the usual forward equations. 

Since the derivation of the forward equation is totally equivalent for both instances $x_T=\pm 1$, let us focus here on the  case $x_T = 1$, \emph{i.e.} on $P_+(x,t|x_0,t_0)$. Our starting point is the Chapman-Kolmogorov equation for Markovian processes~\cite{kampen_stochastic},
\begin{equation}
    P_+(x,t|x_0,t_0) = \int_{-\infty}^1 \!\de x' \, P_+(x,t|x',t') P_+(x',t'|x_0,t_0),
\end{equation}
where $t'$ is any intermediate time,  $t_0\leq t'\leq t$. Note that the upper limit of the integral on the rhs in $x'=1$ due to the presence of a absorbing boundary at that location. Now we differentiate the Chapman-Kolomogorov equation with respect to $t'$,
\begin{eqnarray}
    0 = \int_{-\infty}^1 \!\de x' \, \bigg[&\partial_{t'} P_+(x,t|x',t') P_+(x',t'|x_0,t_0) \nonumber
    \\ &+ P_+(x,t|x',t') \partial_{t'} P_+(x',t'|x_0,t_0)\bigg].
\end{eqnarray}
Introducing the forward equation \eref{eq:forward-adim} for $P_+(x',t'|x_0,t_0)$ and carrying out integration by parts, we obtain
\begin{eqnarray}
    0 = \int_{-\infty}^1 \!\de x' \, \bigg[& \partial_{t'} P_+(x,t|x',t') +\partial_{x'}^2 P_+(x,t|x',t')- \alpha(x')^2 P_+(x,t|x',t')
    \nonumber \\ &  + \alpha(x')^2 P_+(x,t|0,t')\bigg] 
    P_+(x',t'|x_0,t_0).
\end{eqnarray}
Taking the limit $t'\to t_0^+$, we have that $P_+(x',t'|x_0,t_0)\to\delta(x'-x_0)$. Thus, the term in brackets vanishes for $x'=x_0$ and $t'=t_0$, \emph{i.e.} we get the backward equation
\begin{eqnarray}    
-\partial_{t_0} P_+(x,t|x_0,t_0) =  &\partial_{x_0}^2 P_+(x,t|x_0,t_0) 
- \alpha(x_0)^2 P_+(x,t|x_0,t_0) \nonumber \\
&+ \alpha(x_0)^2 P_+(x,t|0,t_0),
\label{eq:backward-adim}
\end{eqnarray}
which has to be solved with the initial condition $P_+(x,t|x_0,t_0=t)=\delta(x-x_0)$. In order to write the backward equation in the form presented in \eref{eq:FPT-dist},  we have only to take into account that the process is homogeneous in time, $P_+(x,t|x_0,t_0)=P_+(x,t-t_0|x_0,0)$ depends on time through the difference $t-t_0$. As discussed in the main text,  the survival probability $S_\pm (t|x_0)$ and the first passage time distribution $f_\pm (t|x_0)$ fulfill exactly the same backward equation. To get them, it suffices to use the definitions in \eref{eq:def_surv} and \eref{eq:def_FPTdist}, which are linear in $P_\pm(x,t|x_0,t_0)$.

\section{Detailed solution to equation~\eref{eq:GMF}}
\label{ap:Lap-sol}
In this section, a detailed solution of equation~\eref{eq:GMF} for $\phi_+$ is provided, submitted to the corresponding boundary conditions that stem from equations~\eref{eq:bc-FPT}:
\begin{eqnarray}
      &\phi_+(s|1)&=1, \label{eq:ap-bc-FPT1}\\
      \lim_{x\to -\infty} \partial_{x_0} &\phi_+ (s|x_0)&=0. \label{eq:ap-bc-FPT2}
\end{eqnarray}
Since the resetting rate takes different values at each side of $x_r=0$, it is handy to find the solution on each side and afterwards match them---enforcing consistency thereof: 
\begin{equation}
    \phi_+(s|x_0) =   \left\{ \begin{array}{@{\kern2.5pt}ll}
    \hfill \phi^{(R)}_+(s|x_0), & \mbox{if }  x_0 \geq 0,\\
    \hfill \phi^{(L)}_+(s|x_0), & \mbox{if }  x_0 \leq 0.
\end{array}
\right. 
\end{equation}
On each side, an inhomogeneous second order differential equation with constant coefficients holds. The general solution is 
\begin{eqnarray}
    \phi_+^{(R)}(s|x_0) = A e^{\lambda_+ x_0} + B e^{-\lambda_+ x_0} + \frac{\alpha_+^2}{\lambda_+^2}\phi_+(s|0)  , \\
    \phi_+^{(L)}(s|x_0) = C e^{\lambda_- x_0}  + \frac{\alpha_-^2}{\lambda_-^2}\phi_+(s|0)  ,
\end{eqnarray}
where we have defined $\lambda_{\pm}\equiv\sqrt{s+\alpha^2_{\pm}}$ and brought to bear equation~\eref{eq:ap-bc-FPT2}.

In the following, $\{A, B , C, \phi_+(s|0) \}$, which are functions of $s$, are determined by imposing (four) appropriate conditions. These are: (i) the boundary equation in equation~\eref{eq:ap-bc-FPT1}, which now reads
\begin{equation}
\label{eq:ap-sys1}
        1 = A e^{\lambda_+  } + B e^{-\lambda_+  } + \frac{\alpha_+^2}{\lambda_+^2}\phi_+(s|0),
\end{equation}
(ii)-(iii) the continuity of $\phi_+(s|x_0)$ at $x_0=0$,
\begin{eqnarray}
\label{eq:ap-sys2}
    \phi_+(s|0) = A  + B + \frac{\alpha_+^2}{\lambda_+^2}\phi_+(s|0), \qquad
    \phi_+(s|0) = C  + \frac{\alpha_-^2}{\lambda_-^2}\phi_+(s|0),
\end{eqnarray}
and (iv) the continuity of the first derivative,  $\partial_{x_0}\phi_+(s|x_0)$, at $x_0=0$,
\begin{equation}
\label{eq:ap-sys4}
    \lambda_- C = \lambda_+ (A-B).
\end{equation}
Insertion of the solution the linear system \eref{eq:ap-sys1}-\eref{eq:ap-sys4} into $\phi_+^{(R,L)}(s|x_0)$ yields
\begin{eqnarray}
\label{eq:ap-sol-GMF1}
    \phi_+^{(R)}(s|x_0) =  \phi_+(s|0) + s\frac{\lambda_{-}\left[\cosh\left(\lambda_{+}x_{0}\right)-1\right]+\lambda_{+}\sinh\left(\lambda_{+}x_{0}\right)}{\lambda_{-}\left(\alpha_{+}^{2}+s\cosh\lambda_{+}\right)+\lambda_{+}s\sinh\lambda_{+}}, \\
\label{eq:ap-sol-GMF2}
    \phi_+^{(L)}(s|x_0) = \phi_+(s|0) + \lambda_{+}^{2} s
    \frac{ e^{\lambda_{-}x_{0}}-1}{\lambda_{-}^{2}\left(\alpha_{+}^{2}+s\cosh\lambda_{+}\right)+s\lambda_{+}\lambda_{-}\sinh\lambda_{+}} ,
\end{eqnarray}
where
\begin{eqnarray}
\label{eq:ap-sol-GMF3}
    \phi_+(s|0)=\frac{\lambda_{+}^{2}\lambda_{-}}{\lambda_{-}\left(\alpha_{+}^{2}+s\cosh\lambda_{+}\right)+s\lambda_{+}\sinh\lambda_{+}} .
\end{eqnarray}
Equation~\eref{eq:GMF-sol} of the main text provides a convenient simplification of $\phi_+(s|0)$, obtained from equation~\eref{eq:ap-sol-GMF3} after substituting $\lambda_{\pm}=\sqrt{s+\alpha^2_{\pm}}$.

\section{The unique real pole for the first passage time distribution}
\label{ap:uni-pole}

The long-time behaviour of a function $m(t)$ can be derived from its Laplace transform $\mu(s)$. Assuming that the singularity of $\mu$ with the highest real part $s^*$ is a simple pole, 
\begin{equation}
    \label{eq:Gen-Asymp}
    m(t) = \frac{1}{2\pi i}\int_{\sigma_-i\infty}^{\sigma+i\infty} 
    ds \, \mu(s)\,e^{st} \sim \mbox{Res}[\mu(s)e^{s^* t};s^*], \; t\to\infty.
\end{equation}
as a consequence of  Cauchy's residue theorem. 

\begin{figure}
    \centering
\includegraphics[width=0.7\textwidth]{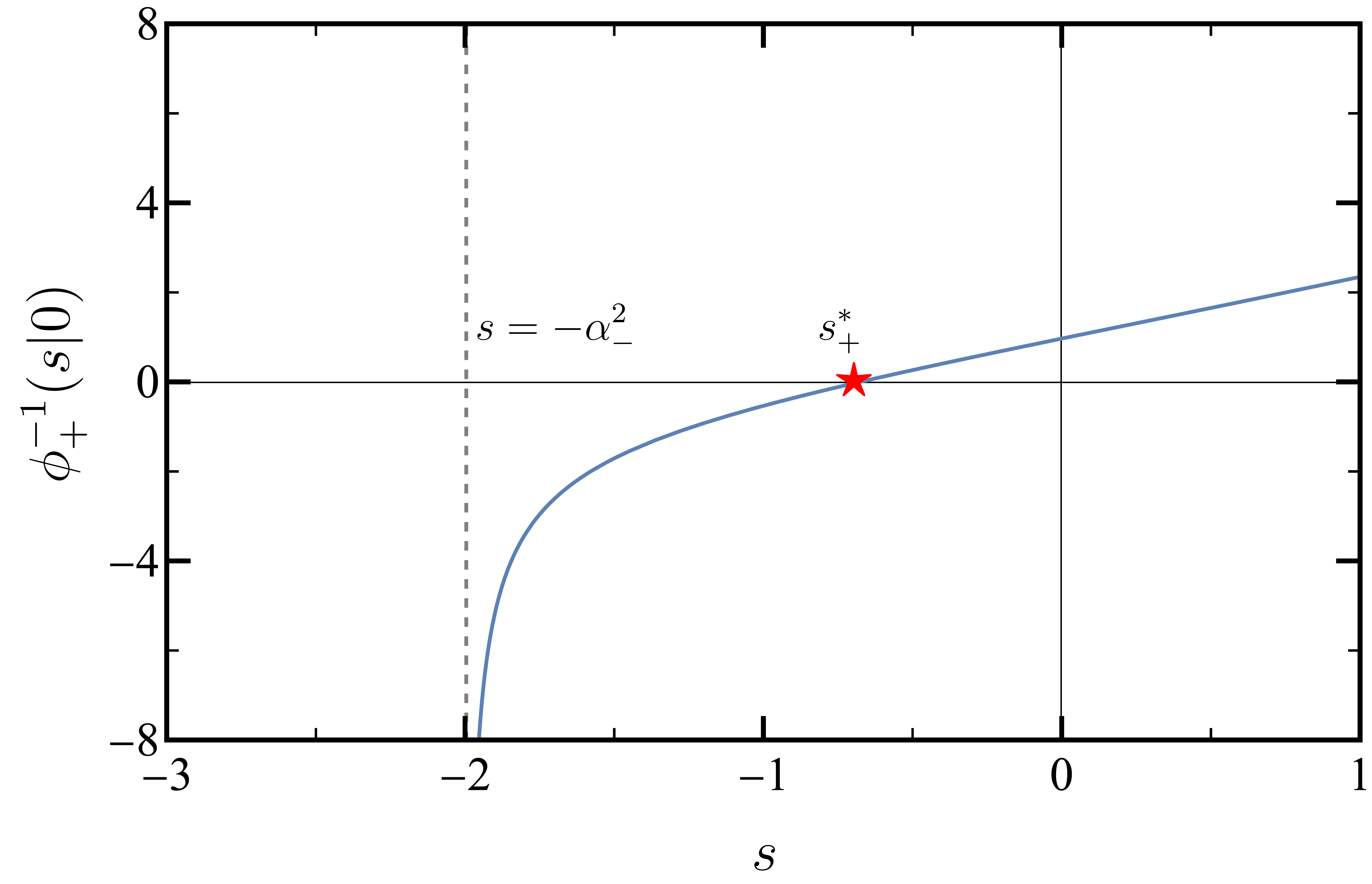}
    \caption{Inverse of the Laplace transform of the moment generating function over the real axis. The dashed line corresponds to the vertical asymptote at $s=-\alpha_-^2$, whereas the red star marks the position of the simple pole $s_+^*$.}
    \label{fig:real_Pole}
\end{figure}

Herein, we derive the asymptotic expression for the moment generating function provided in the main text, and show that it depends on the largest real pole $s_+^*$ of $\phi_+(s|0)$.  The poles of the moment generating function in equation~\eref{eq:GMF-sol} can be computed analysing when $\phi_+^{-1}(s|0)$ vanishes. First, we focus on the real axis,  where we can prove that there is always a unique negative real pole. For any $(\alpha_+$, $\alpha_-)$, we have that
\begin{eqnarray}
    \lim_{s \to (-\alpha_-^{2})^+} \phi_+^{-1}(s|0) =-\infty, \quad
    \phi_+^{-1}(0|0) = 1, \quad 
    \lim_{s \to \infty} \phi_+^{-1}(s|0) =\infty,
\end{eqnarray}
as shown in  figure~\ref{fig:real_Pole} for $(\alpha_+=1$, $\alpha_-=\sqrt{2})$. Furthermore, since its first derivative is always positive, $\partial_s \phi_+^{-1}(s|0)>0$,  just one simple pole $s^*_+$ exists in the interval $(-\alpha_-^2,0)$. Note that, despite $s=-\alpha_-^2$ is a branching point, the real pole is always larger than it.

We have shown the existence of a real pole higher than the branching point of the moment generating function. Nevertheless, it is necessary to check that this is indeed the pole with the highest real part in the whole complex plane. We do not have an analytical proof for this, but the numerical evaluation of the function gives no other pole with higher real part. Once we have shown that $s^*_+$ corresponds to a simple pole, it is fully justified to use  equation~\eref{eq:Gen-Asymp}, which leads to equation~\eref{eq:asymp+1} in the main text. Due to the symmetry of the problem, which we have repeatedly made use along this work, a similar result applies to $\phi_-(s|0)$, with its largest real pole $s_-^*$.

\section{Exponential rates for the minimum mean first passage time}
\label{ap:opt-poles}

The excellent agreement between simulations and analytical approximations have been shown in figure~\ref{fig:crossover} of the main text. In the theoretical prediction, the poles from $\phi_+(s|0)$ and $\phi_-(s|0)$, which depend on the set of parameters $(\alpha_+,\alpha_-,p)$, play a central role. 

\begin{figure}
    \centering
\includegraphics[width=0.7\textwidth]{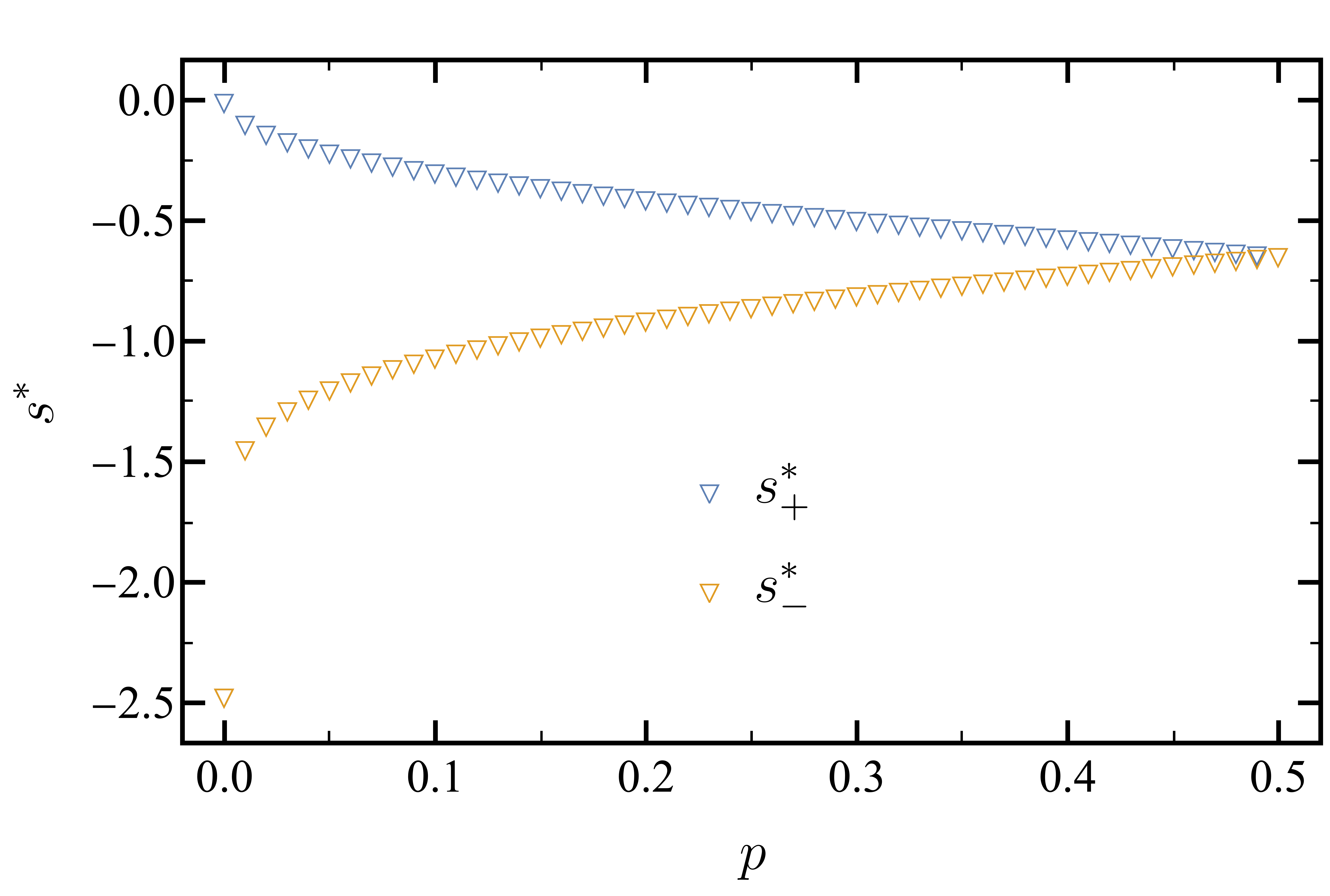}
    \caption{Poles $s_{\pm}^*$ as a function of the disorder $p$. The remainder of the model parameters, \emph{i.e.}~the resetting rates $(\alpha_+,\alpha_-)$ have been chosen as those minimising the MFPT---consistently with figure~\ref{fig:crossover} in the main text.}
    \label{fig:Optimal_Poles}
\end{figure}

In this appendix, we complete the picture by displaying in figure~\ref{fig:Optimal_Poles} the dependence of the poles on $p$ for the same choice of $(\alpha_+,\alpha_-)$ in figure~\ref{fig:crossover}, \emph{i.e.}~that minimising the MFPT derived in section~\ref{sec:opt}. For $0<p< 1/2$, it is always $s_+^*>s_-^*$ and thus the dominant contribution for long times comes from $s_+^*$. Both poles get closer as $p$ increases and merge for $p=1/2$, consistently with the behaviour shown in the last panel in figure~\ref{fig:crossover}. In the limit as $p\to 0^+$, $s_+^*$ tends to zero: it is from this behaviour that there emerges the slow decay of $f_+(t|0)$---as compared with that for $p=0$---shown in the second panel of figure~\ref{fig:crossover}. The behaviour of $f(t|0)$ for $p=0^+$ is thus very different from that for $p=0$, shown in the first panel, where $f_-(t|0)$ is the only survivor. We recall that the optimal resetting rates at $p=0$ (the target is certainly to the left of the resetting point) are $\alpha_+ \to \infty$, $\alpha_-=0$. This means that the moment generating function reduces to $\phi(s|0)=\phi_-(s|0)=\sech(\sqrt{s})$, whose pole with the largest real part is $s_{-}^*=-2.47$.

\section{Asymptotic behaviour for optimal first passage time}
\label{ap:asymp-MFPT}

In this appendix, we provide the explicit derivation of the behaviour of the optimal resetting rates in the vicinity of $p=0$ (no uncertainty about the target location) and $p=1/2$ (maximum uncertainty about the target location). The theoretical predictions stemming from this analysis have been plotted in figure~\ref{fig:minMFPT} of the main text.

\subsection*{Maximum uncertainty about the target location}

When the dichotomous target distribution is almost symmetric, \emph{i.e.}~$p = 1/2+\delta p$ with $\delta p \ll 1$, it is handy to rewrite the MFPT in equation~\eref{eq:MFPT1} as
\begin{equation}\label{eq:F-symm}
\tau^{(1)} =  F_s(\alpha_+,\alpha_-) + F_a(\alpha_+,\alpha_-) \delta p,
\end{equation}
where the functions 
\begin{eqnarray}
F_s(\alpha_+,\alpha_-)=\frac{F_1(\alpha_+,\alpha_-)+F_1(\alpha_-,\alpha_+)}{2},  \label{eq:ap-Fs}\\
F_a(\alpha_+,\alpha_-)=F_1(\alpha_+,\alpha_-)-F_1(\alpha_-,\alpha_+), \label{eq:ap-Fa}
\end{eqnarray}
are symmetric and antisymmetric under the exchange $\alpha_+\leftrightarrow\alpha_-$. From homogeneous resetting, it is known that $F_s$ is optimal for $\alpha_+=\alpha_-=\widetilde{\alpha}_{\symm}$, where the symmetric optimal rate $\widetilde{\alpha}_{\symm}$ is given by the equation 
\begin{equation}
    \label{eq:alpha-sym}
    \widetilde{\alpha}_{\symm}= 2\left(1- e^{-\widetilde{\alpha}_{\symm}}\right) \Rightarrow \widetilde{\alpha}_{\symm} = 2 + W_0(-2 e^{-2}) 
    ,
\end{equation}
where $W_0$ is the principal branch of the Lambert $W$ function \cite{Evans11prl,Evans11jphysa}.

For $\delta p\ll 1$, we write $(\alpha_+,\alpha_-)= (\widetilde{\alpha}_{\symm},\widetilde{\alpha}_{\symm}) + (c_+,c_-) \delta p$ and the MFPT reads
\begin{equation}
\tau^{(1)} \simeq F_s + {\bf c}^T \cdot {\nabla F_s} \, \delta p + \frac{1}{2} {\bf c}^T \cdot \mathbb{H}_s \cdot {\bf c} \, (\delta p)^2 + F_a \delta p + {\bf c}^T \cdot {\nabla F_a} \, (\delta p)^2
\end{equation}
where ${\bf c}^T\equiv(c_+,c_-)$, $\mathbb{H}_s$ is the Hessian matrix of $F_s$. All the coefficients 
are computed at $\alpha_+=\alpha_-=\widetilde{\alpha}_{\symm}$, thus $\nabla F_s=0$ because of optimality and $F_a=0$ because of antisymmetry.  Taking derivative with respect to $c_\pm$ and equating to zero, we get
\begin{equation}
\label{eq:c_opt}
\widetilde{{\bf c}} = - \mathbb{H}_s^{-1} \cdot \nabla F_a,
\end{equation}
for the optimal value of $\bf{c}$. The minimum MFPT is then
\begin{equation}
\widetilde{\tau}^{(1)} \simeq F_s - \frac{1}{2} \left({\nabla F_a}\right)^T \cdot \mathbb{H}^{-1}_s \cdot \nabla F_a (\delta p)^2 \leq F_s \ .
\label{eq:tau-near12}
\end{equation}
Making use of equation~\eref{eq:alpha-sym}, we would be able to get the specific values for the coefficients,
\begin{eqnarray}
        \left( \nabla F_a \right)^T= \frac{4-\widetilde{\alpha}_{\symm}}{2\widetilde{\alpha}_{\symm}^2(2-\widetilde{\alpha}_{\symm})} (1,-1) 
        ,
        \\
   \mathbb{H}_s = \frac1{4\widetilde{\alpha}_{\symm}^3(2-\widetilde{\alpha}_{\symm})} 
\left(\begin{array}{cc}
\widetilde{\alpha}_{s}(1+\widetilde{\alpha}_{s}) & -4+\widetilde{\alpha}_{s}(3-\widetilde{\alpha}_{s})\\
-4+\widetilde{\alpha}_{s}(3-\widetilde{\alpha}_{s}) & \widetilde{\alpha}_{s}(1+\widetilde{\alpha}_{s})
\end{array}\right) 
,
\\
    \widetilde{\bf c}^T = -\frac{\widetilde{\alpha}_{\symm}(4-\widetilde{\alpha}_{\symm})}{2+\widetilde{\alpha}_{\symm}(\widetilde{\alpha}_{\symm}-1)} (1,-1)
    ,
    \\
    \frac{1}{2} \left({\nabla F_a}\right)^T \cdot \mathbb{H}_s^{-1} \cdot \nabla F_a = \frac{(4-\widetilde{\alpha}_{\symm})^2}{2\widetilde{\alpha}_{\symm} (2-\widetilde{\alpha}_{\symm})(2+\widetilde{\alpha}_{\symm}(\widetilde{\alpha}_{\symm}-1))} 
    .
    \label{eq:tau-near12-coeffs}
\end{eqnarray}

We have cut the expansion for $(\widetilde{\alpha}_+,\widetilde{\alpha}_-)$ above at linear terms, which provides good estimates in the vicinity of $p=1/2$. Nonetheless, higher-order corrections can be incorporated to the expansion in a systematic manner, without any technical difficulty.

\subsection*{Minimum uncertainty about the target location}

In the opposite regime, $p = \delta p \ll 1$ (recall that we have restricted ourselves to $0\le p\le 1/2$ because of the symmetry of the problem), it is almost sure that the target lies on the half-line to the left of the resetting point. As physically expected, for $p=0$, 
the optimal rates are $\alpha_-=0$ and $\alpha_+ \to \infty$, \emph{i.e.} a reflecting boundary at the origin and free diffusion for any $x<0$. Therefore, one expects to have  $\alpha_-\ll 1$ and $\alpha_+\gg 1$ in the limit as $p\to 0^+$. 

Our starting point is equation~\eref{eq:MFPT1} fot the MFPT, which now reads
\begin{equation}\label{eq:F-asympt}
\tau^{(1)} = F_1(\alpha_-,\alpha_+) + \delta p \left[ F_1(\alpha_+,\alpha_-) - F_1(\alpha_-,\alpha_+) \right].
\end{equation}
In the limit $\alpha_-\ll 1$ and $\alpha_+\gg 1$, the leading terms in $F_1$ are
\begin{eqnarray}  
    F_1(\alpha_+,\alpha_-) = \frac{e^{\alpha_+}}{2\alpha_+ \alpha_-} + o \left( \frac{e^{\alpha_+}}{\alpha_+ \alpha_-}  \right), \\
    F_1(\alpha_-,\alpha_+) = \frac{1}{2} + \frac{\alpha_-^2}{24} + \frac1{\alpha_+} + o\left(\alpha_-^2\right) + o\left(\alpha_+^{-1} \right).
\end{eqnarray}
To leading order, this entails that 
\begin{eqnarray}
    \frac{\partial \tau^{(1)}}{\partial \alpha_+} \sim - \frac1{\alpha_+^2} + \frac{\alpha_+-1}{2\alpha_+^2 \alpha_-} e^{\alpha_+} \delta p \sim - \frac1{\alpha_+^2} + \frac{e^{\alpha_+}}{2\alpha_+ \alpha_-} \delta p, \\
\frac{\partial \tau^{(1)}}{\partial \alpha_-} \sim \frac{\alpha_-}{12} - \frac{e^{\alpha_+}}{2\alpha_+ \alpha_-^2} \delta p.
\end{eqnarray}
For obtaining the optimal values $(\widetilde{\alpha}_+,\widetilde{\alpha}_-)$, we enforce ${\partial \tau^{(1)}}/{\partial \alpha_{\pm}}$ to be zero, which gives us the leading terms of the optimal resetting rates for $\delta p\ll 1$, 
\begin{eqnarray}
    \widetilde{\alpha}_+\sim 2 W_0 \left( \frac{3^{1/4}}{\sqrt{\delta p}}\right), \quad 
    \widetilde{\alpha}_-\sim \frac{2\sqrt{3}}{\widetilde{\alpha}_+}.
    \label{eq:alpha-near0}
\end{eqnarray}

\begin{figure}
    \centering
    \includegraphics[width=0.7\textwidth]{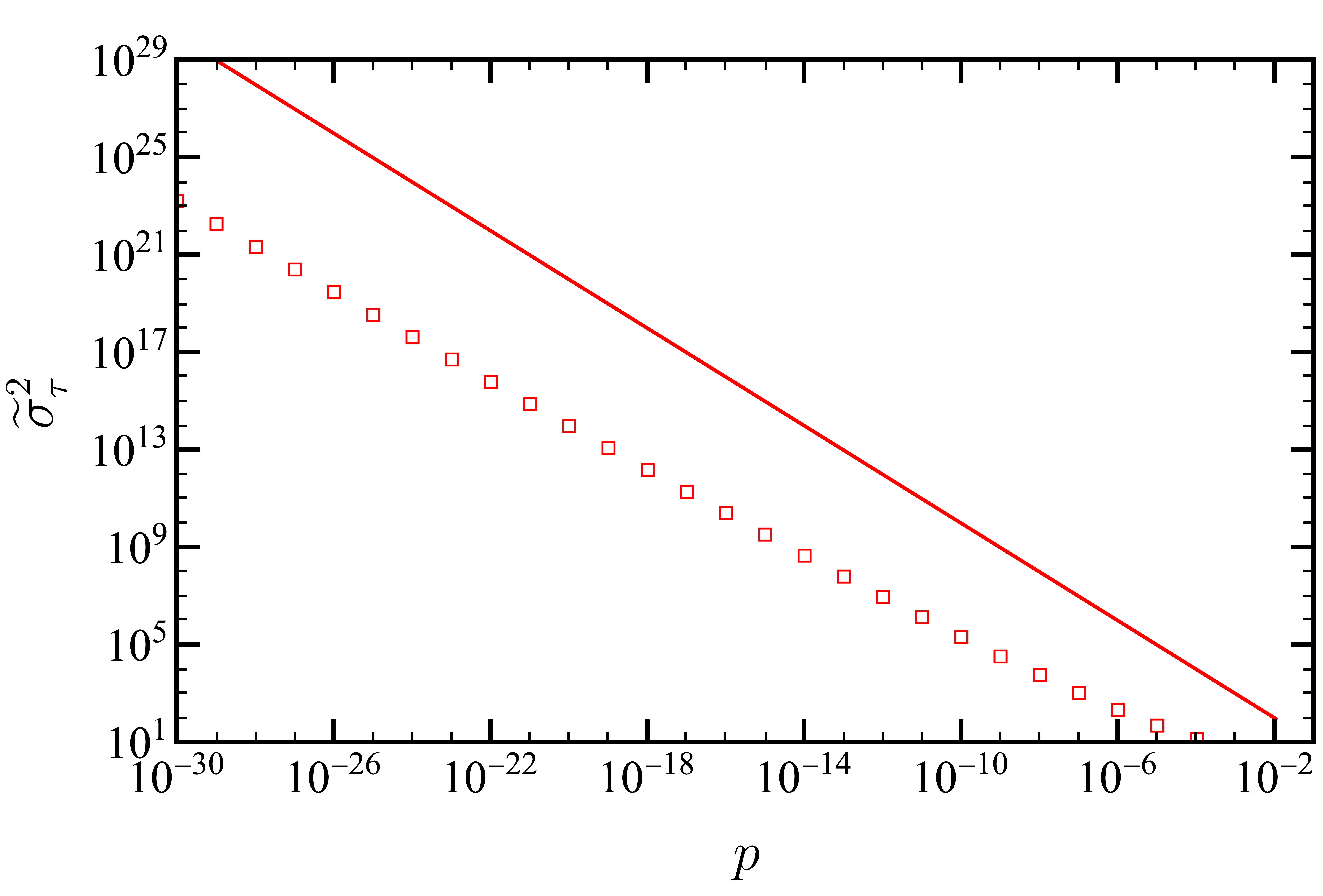}
    \caption{Divergence of the variance at minimum MFPT in the limit as $p \to 0^+$. The numerical solutions (symbols) are compared with the leading order of the asymptotic prediction $p^{-1}$ (solid line).
    }
    \label{fig:div-second}
\end{figure}

It is interesting to estimate the value of the pole $s_+^*$ in this limit, for the optimal rates just derived above. It is the vanishing of $s_+^*$ in the limit as $p\to 0^+$ that explains the divergence of the second moment---and thus of the standard deviation---therein. When looking for the poles of the moment generating function~\eref{eq:GMF-sol}, a dominant balance argument leads to the asymptotic behaviour $s_+^*\sim -4\sqrt{3}e^{-\alpha_+}$; note that $|s_+^*|< \widetilde{\alpha}_-^2$ as it should be, in fact $|s_+^*|\ll \widetilde{\alpha}_-^2\propto \widetilde{\alpha}_+^{-2}$. Taking into account~\eref{eq:alpha-near0}, one has that $s_+^*\sim -4 \sqrt{3} p (\ln p)^2$: the contribution of this pole to the MFPT is proportional to $p/s_+^*$ and therefore vanishes, whereas its contribution to the second moment, which is proportional to $p/(s^*_+)^2$, diverges as $p^{-1}$ (with slowly varying logarithmic corrections) as shown in figure~\ref{fig:div-second}.


\section*{References}
\bibliographystyle{iopart-num}
\bibliography{disorderedreset}

\end{document}